\documentclass[aps,showpacs,a4paper,floatfix,twocolumn,prc,amsmath,amssymb]{revtex4-1}
\usepackage{graphicx}
\usepackage{epsfig}
\usepackage{color} 
\usepackage{morefloats}
\usepackage{rotating}
\usepackage{relsize}
\usepackage{url}
\usepackage{natbib}
\usepackage[breaklinks=true]{hyperref}

\usepackage{booktabs}
\usepackage{soul}
\usepackage{array}

\begin{document}

\title{Light hyperclusters and hyperons in low-density hot stellar matter}

\author{Tiago Cust{\'o}dio}
\author{Helena Pais}
\author{Constan\c ca Provid{\^e}ncia}

\affiliation{CFisUC, Department of Physics, University of Coimbra,
	3004-516 Coimbra, Portugal.}

\begin{abstract}
	The abundance of light  nuclei and
	hyperons, that are produced in stellar environments such as
	supernova or binary mergers, is calculated
	within a relativistic mean-field model with density dependent
	couplings in low-density matter. Five light nuclei 
	are considered, together
	with three light hypernuclei.
	We show that the presence of hyperons shifts the dissolution of
	clusters to larger densities, and increases the amount of clusters. This effect is larger the smaller the
	charge fraction, and the higher the temperature.  The abundance of hyperons is also
	affected by the cluster formation: neutral and positively charged
	hyperons suffer a reduction, and the negatively charged ones an
	increase. 
	We also observe that the
	dissolution of the less-abundant clusters occurs at larger densities
	due to smaller Pauli-blocking effects. Overall, hypernuclei set in at
	temperatures above 25 MeV, and depending on the temperature and chemical
	composition, they may be more abundant than $\alpha$-particles, or even more abundant than other heavier clusters.  
\end{abstract}

\maketitle

\section{Introduction}

Light nuclei are found in core-collapse supernova matter and in binary
neutron star (NS) mergers. Their presence may impact the evolution of
these systems by affecting the rate at which the weak reactions take place
during the core collapse \cite{Arcones2008,Fischer2020},  or the dissolution  of the remnant torus of accreted matter that is formed around the high mass NS after a binary merger
\cite{Rosswog2015}. Light clusters could also influence the dissipative
processes that determine the post-merger evolution and mass ejection
from the remnant \cite{Alford2018,Fujibayashi2017}.

These clusters have been detected in heavy ion collisions in several experiments, such as ALICE at the Large Hadron Collider (LHC), STAR at the Relativistic Heavy Ion Collider (RHIC), or J-PARC, from the E13 collaboration.  Some of these states, like the
deuteron, the hypertriton  \cite{Adam2019}, the  hyper-hidrogen4
\cite{Esser2014} or the hyperhelium4
\cite{Yamamoto2015} are loosely bound
objects with quite a large radius. It is still not understood why
these states are well described within a thermal approach with a
temperature production of the order of 150 MeV, much larger than their
binding energy \cite{Braun-Munzinger2018}. At RHIC and LHC, the baryonic 
chemical potential is quite low. The formation of light clusters at
much smaller temperatures, of the order of 5 to 12 MeV, but larger
densities, below 0.1 fm$^{-3}$, has been measured by the multi-detectors
NIMROD at the Texas A\&M University  \cite{Qin2011} and  INDRA \cite{indra}
at GANIL. These last measurements can help understand the low-density
nuclear matter equation of
state (EoS) at temperatures and densities of interest to
the evolution of supernovae and binary neutron star mergers.

Neutron stars are formed by cold catalysed $\beta$-equilibrium matter  constituted by neutrons,
protons, electrons and muons below a density $\approx
2n_0-3n_0$, where $n_0\approx 0.15$ fm$^{-3}$ is the
saturation density of symmetric nuclear matter. At
larger densities other degrees of freedom such as hyperons, deltas,
kaon or pion condensates or quark matter may set in
\cite{Glend}. During the supernova or binary merger evolution,
$\beta$-equilibrium is not necessarily achieved and temperatures as
high as 50 to 100 MeV may be attained. To describe these events, it is
necessary, therefore, to consider a wide range of electron fractions, temperatures
and densities. In Refs.~\cite{Marques2017,Fortin2017}, it has been shown that the
inclusion of the complete baryonic octet decreases the free energy of
matter, and the EoS based on relativistic mean-field models, like DD2
\cite{Typel2009} or SFHo \cite{Steiner2012}, including the complete
baryonic octet, have been built and made available in the CompOSE database
\footnote{https://compose.obspm.fr/}. In both of these studies, it
was shown that, at low densities, the hyperons compete with light
nuclear clusters, and the minimization of the free energy should allow
for the appearance of hyperons at very low densities,  which, however,
was not implemented.

However, in the nineties, one could already find EoS with light clusters included, like the  general-purpose EoS by Lattimer and Swesty \cite{LS}, or the EoS by Shen {\it et al.} \cite{Shen1998},
based on the single-nucleus approximation. In both cases, light
clusters were restricted to $\alpha-$particles. Improved models of the non-homogeneous
matter at finite temperature containing light clusters in the
framework of nuclear statistical equilibrium were proposed
later \cite{Hempel2010,Raduta2010}. In these models, the
introduction of an excluded volume is necessary to dissolve the
clusters at high densities. A different  approach was undertaken in Refs.~\cite{Typel2009,Avancini2010,Ferreira2012}, where density effects are included  within a
relativistic mean-field  approach that  describes  the light
clusters as new particles which couple to the mesonic fields. The
model, first published in Ref.~\cite{Typel2009}, introduces the
temperature-dependent  cluster binding shifts determined
from a quantum statistical approach to nuclear matter in thermodynamic
equilibrium  \cite{Ropke2008,Ropke2011,Ropke2015,Ropke2020}. This
model was recently improved, by taking into account continuum correlations, and it was
applied to  simulations of core-collapse supernovae \cite{Fischer2020}.  In particular,
the authors analysed the effect of medium modifications on the simulations.

In Ref.~\cite{Menezes2017}, the possible appearance of hyperons in the density
region of the non-homogeneous matter that forms the inner crust of a neutron star was analyzed. Temperatures below the melting temperature of the heavy clusters that
form this region were considered, i.e $T \lesssim 15$ MeV.  It
was found that only very small amounts of hyperons, like $\Lambda$ fractions below
10$^{-5}$, were present in the background gas. The
low-density EoS of stellar matter including light
clusters and heavy baryons was also studied in Ref.~\cite{Sedrakian2020}. In addition to hyperons, the author also considered delta-baryons, pions, and the presence of a representative heavy
cluster. It was shown that, depending on temperature and density, the
composition of matter may shift from a greater abundance of light clusters to a heavy-baryon predominance.

In the present work, we are going to simultaneously calculate, in a consistent way, the abundance of light nuclei and hypernuclei, as well as hyperons, within the DD2 relativistic mean-field (RMF) model
\cite{Typel2009}, taking for the  $\sigma-$meson cluster coupling the
value obtained in \cite{Custodio2020}. The introduction of light clusters is going to follow the approach first
presented in Ref.~\cite{Pais2018}, where the effect of the medium on the
binding energy of the clusters is considered through the introduction
of a binding energy shift, together with a universal coupling of the $\sigma-$meson to the
different light clusters, that was chosen so that the equilibrium constants
of the NIMROD experiment \cite{Qin2011} were reproduced.  In Refs.~\cite{Pais2020prl,Pais2020}, the same
approach was applied to the description of the INDRA data
\cite{indra} including the medium effects on the data analysis. It was
verified that, in this case,  the equilibrium constants could be
reproduced only if a  larger   $\sigma-$meson coupling was
introduced. The calibration of the $\sigma-$meson to the clusters coupling
was later performed for other models in Ref.~\cite{Custodio2020}.


The paper is organized as follows: in  Sec.~\ref{sec2}, we introduce the formalism,  in Sec.~\ref{sec3}
the results are presented for different scenarios: the effect of
temperature, charge fraction, and density, and the inclusion of hyperons,
light clusters and hyperclusters. Finally, in
Sec.~\ref{sec4}, some conclusions  are drawn.

\section{Formalism\label{sec2}}

In this section, we present the model used throughout the
paper, and we discuss how hyperons, light clusters and light hyperclusters, which are considered as
point-like particles, are included within our approach.

Our system's gas is constituted by unbound neutrons ($ n $) and protons ($p$), as well the following six hyperons: $ \Lambda $, $\Sigma^-$, $\Sigma^0$, $\Sigma^+$, $\Xi^-$, $\Xi^0$. Together, these eight particles form the spin-1/2 baryonic octet. 

Immersed in this gas, we will also consider five purely nucleonic light nuclei ($^{2}$H, $^3$H, $^3$He, $^4$He, $^6$He) as well as three hypernuclei: the $^{3}_{\Lambda}\text{H}$ hypertrition, the $^{4}_{\Lambda}\text{H}$ hyper-hidrogen4 and the hyperhelium4 $^{4}_{\Lambda}\text{He}$. For each of these three hypernuclei, a nucleon was replaced by a $\Lambda$ hyperon. In Table \ref{table_spin_isopin}, the spin and isospin projection quantum numbers can be found for each particle considered here.

\begin{table*}[htb]
\caption{\label{table_spin_isopin}
Spin ($ J $) and isospin projection ($ I_3 $) quantum numbers for all particles considered in our system.}
\begin{ruledtabular}
\begin{tabular}{ccccccccccccccccc} 
		& $ n $    & $ p $ & $ \Lambda $ & $ \Sigma^-$ & $ \Sigma^0 $ & $ \Sigma^+ $ & $ \Xi^- $  & $ \Xi^0 $ & ${}^{2}\text{H}$ & ${}^{3}\text{H}$   & ${}^{3}\text{He}$ & ${}^{4}\text{He}$ & ${}^{6}\text{He}$ & $^{3}_{\Lambda}\text{H}$ &$^{4}_{\Lambda}\text{H}$  & $^{4}_{\Lambda}\text{He}$ \\ 
		\hline 
		$J $   & $1/2$    & $1/2$ & $1/2$      & $1/2$      & $1/2$      & $1/2$       & $1/2$     & $1/2$   & $ 1 $  & $1/2$   & $1/2$    & $0$   & $0$   &$1/2$   & $0$    & $0$    \\
		$ I_3 $ & $-1/2$  & $1/2$  & $0$      & $-1$     & $0$      & $1$      & $-1/2$  & $1/2$    & $0$  & $-1/2$  & $1/2$   & $0$   & $-1$  & $0$   & $-1/2$  & $1/2$ \\ 
\end{tabular}
\end{ruledtabular}
\end{table*}

In the RMF theory, the interactions between different baryons are mediated by the exchange of virtual mesons. Here we will consider the following four mesons: the isoscalar-scalar $\sigma$ meson field that provides the attractive strong force; the isoscalar-vector $\omega^{\mu}$ meson field responsible for the repulsive strong force; the isoscalar-vector $\phi^{\mu}$ meson field responsible for an extra repulsion between two hyperons, and the isovector-vector $\vec{\rho^{\mu}}$ meson field which accounts for the isospin dependence of the interactions. The baryons are described by a Dirac spinor.

The Lagrangian density for this system reads \cite{Glend,Typel2009,Fortin2017,Pais2018}:

\begin{equation}\label{Lagrangian_nlm_inicial}
	\mathcal{L}=\!\sum_{\substack{b=baryonic \\ octet}} \!\!\!\!\!\!\mathcal{L}_b + \!\!\sum_{\substack{i=light \\ nuclei}} \!\!\!\mathcal{L}_{i} + \!\!\!\!\sum_{\substack{j=light \\ hypernuclei}} \!\!\!\!\!\!\mathcal{L}_{j} + \!\!\sum_{\substack{m=\sigma,\omega,\phi,\rho }}\!\!\!\!\mathcal{L}_m.
\end{equation}
The subscript $ b $ stands for the eight particles belonging to the spin-1/2 baryonic octet ($ n,p,\Lambda,\Sigma^-,\Sigma^0,\Sigma^+,\Xi^-,\Xi^0 $). 

\subsection{The homogeneous gas}

The Lagrangian density of the gas, which includes the spin-1/2 baryon octet, is given by 
\begin{eqnarray}
	\mathcal{L}_b= \bar{\Psi}_b & \left[ i \gamma_{\mu} \partial^{\mu} - m_b + g_{\sigma b} \sigma - g_{\omega b} \gamma_{\mu} \omega^{\mu} \right. \\ & \left. - g_{\rho b} \gamma_{\mu} \vec{I}_b\cdot \vec{\rho^{\mu}} - g_{\phi b} \gamma_{\mu} \phi^{\mu} \right]  \Psi_b \, , \nonumber
\end{eqnarray}
with $\Psi_b$ the baryon field, and $ \vec{I}_b $ the isospin operator. The quantities $ g_{mb} $ 
are the coupling constants of the interactions between the baryons and the mesons. 

We take for the vacuum proton and neutron mass an average value, $ m=m_n=m_p= $ 939 MeV. For the hyperons, we consider the following masses: $ m_{\Lambda}=1115.683 $ MeV, $ m_{\Sigma^-}=1197 $  MeV, $ m_{\Sigma^0}=1193 $ MeV, $ m_{\Sigma^+}=1189 $ MeV, $ m_{\Xi^-}=1321 $ MeV, and $ m_{\Xi^0}=1315 $ MeV.

The coupling constants $ g_{mN} $ of the nucleons ($ N=n,p $) to the $ \sigma$, $\omega$ and $\rho$ mesons are given by the RMF model DD2 \cite{Typel2009} with density-dependent coupling constants. These couplings are written in the form
\begin{equation}
  g_{mN}(n_B) = g_{mN}(n_0) h_M(x)~,\quad x = n_B/n_0~,
\end{equation}
where the  density $n_B$ is the baryonic density. For the isoscalar couplings, the function $h_M$ is given by ~\cite{Typel2009},
\begin{equation}
h_M(x) = a_M \frac{1 + b_M ( x + d_M)^2}{1 + c_M (x + d_M)^2}
\end{equation}
and for the isovector couplings has the form
\begin{equation}
h_M(x) = \exp[-a_M (x-1)] ~.
\end{equation}
The values of
the parameters $a_M, b_M, c_M,$ and $d_M$ are given in Ref.~\cite{Typel2009}. \\

As for the hyperons ($ \Lambda,\Sigma^-,\Sigma^0,\Sigma^+,\Xi^-,\Xi^0 $), their coupling constants $ g_{mb} $ can be defined in terms of the nucleon couplings as $ g_{mb}=R_{mb} g_{mN} $, for $ m=\sigma,\omega,\rho $  and  $ g_{\phi b}=R_{\phi b} g_{\omega N} $.  All these coupling constants of the mesons to the different hyperons, normalized to the respective meson nucleon coupling, can be found in Table~\ref{table_hyperon_couplings}.
In the case of the $\phi-$meson, the coupling to the $\omega$ meson is used instead, because the nucleons do not couple to this meson (their coupling is zero, $ g_{\phi N}=0 $). This is the ``ideal mixing'' scenario \cite{Weissenborn2011}, where the hyperon couplings to the $ \omega $ and $\phi$ mesons is fixed using the SU(6) quark model.  For the hyperon couplings to the $\rho$ meson, they are the same as for the nucleons. Concerning this coupling, what differentiates each hyperon is simply their isospin projection. This will also be true for all the clusters and hyperclusters considered in our model.

The coupling of the $\Lambda$ hyperon to the $\sigma$ meson can be calibrated by fitting the experimental binding energy of $\Lambda$ hypernuclei as described in \cite{Fortin2017}. From Ref.~\cite{Fortin2017}, we chose to use the value of the DDME2D-a model, $ R_{\sigma \Lambda}=0.621 $. Similarly, for the $\Xi$ coupling to the $\sigma$ meson, we use the calibrated value obtained in there, $ R_{\sigma \Xi}=0.320 $ \cite{Fortin2020}. Finally, according to Ref.~\cite{Gal2016}, the $\Sigma$ potential in symmetric nuclear matter lies in the range  $ U_{\Sigma}^{(N)}(n_0)\approx30\pm20 $ MeV, so we fix $ U_{\Sigma}^{(N)}(n_0)=30 $ MeV, and we obtain $ g_{\sigma \Sigma}=0.474 $.

\begin{table}[htb]
\caption{\label{table_hyperon_couplings}
Coupling constants of the mesons to the different hyperons, normalized
to the respective meson nucleon coupling, i.e. $R_{Mj}=g_{Mj}/g_{MN}$,
except for the $\phi-$meson. Here the $g_{\omega N}$ is used for normalisation.}
\begin{ruledtabular}
\begin{tabular}{ccccc}
	$ b $	& $ R_{\sigma b} $   &$ R_{\omega b} $   & $ R_{\phi b} $            & $ R_{\rho b} $ \\ 
	\hline
		$ \Lambda $ & 0.621 & 2/3 & $ -\sqrt{2}/3 $  & 1   \\
	    $ \Sigma $ & 0.474 & 2/3 & $-\sqrt{2}/3 $  & 1   \\ 
		$ \Xi $ & 0.320 & 1/3 &$ -2\sqrt{2}/3 $  & 1  \\ 
	\end{tabular}
	\end{ruledtabular}
\end{table}

\subsection{The light clusters}

Following Refs.~\cite{Pais2018,Pais2020prl}, the Lagrangian density for the fermionic spin-1/2 light nuclei reads 
\begin{equation}\label{key}
	\mathcal{L}_{i}=\bar{\Psi}_{i}\left[\gamma_{\mu}i D_{i}^{\mu} - M_{i}^* \right]\Psi_{i}, \hspace{0.1cm}i={}^{3}\text{H},{}^{3}\text{He} \, ,
\end{equation} 
and for the bosonic light nuclei 
is given according to their spins, 
\begin{eqnarray}
	\mathcal{L}_i&=& \frac{1}{2}\left( iD_{i}^{\mu} \Psi_{i} \right)^*\left( iD_{\mu i} \Psi_{i} \right)  \\ && - \frac{1}{2}\Psi_{i}^*(M_i^*)^2\Psi_{i},\hspace{0.1cm}i={}^{4}\text{He},{}^{6}\text{He} \, ,  \nonumber \\ 
		\mathcal{L}_i&=&\frac{1}{4}\left( iD_{i}^{\mu} \Psi_{i}^{\nu}-iD_{i}^{\nu} \Psi_{i}^{\mu} \right)^*\left( iD_{\mu i} \Psi_{\nu i}-iD_{\nu i} \Psi_{\mu i} \right)  \\ && -\frac{1}{2}\Psi_{i}^{\mu*}(M_i^*)^2\Psi_{\mu i},\hspace{0.1cm}i={}^{2}\text{H} \, , \nonumber
\end{eqnarray}
with 
\begin{equation}\label{key2}
	i D_{i}^{\mu}=i \partial^{\mu} - g_{\omega i} \omega^{\mu} - g_{\rho i}\vec{I}_{i}\cdot \vec{\rho^{\mu}} \, .
\end{equation}
$ g_{\omega i} $ and $g_{\rho i}$ are the couplings of cluster $ i $ to the $\omega$ and $\rho$ mesons, respectively.  They are defined as $ g_{\omega i} = A_{i}g_{\omega N} $, with $ A_i $ the cluster mass number, and $g_{\rho i}=g_{\rho N}$. The effective mass of the cluster $i$, $ M_i^* $, is given by 
\begin{equation}\label{effective_mass_clusters}
	M_i^*=A_i m - g_{\sigma i}\sigma - (B_i^0 +\delta B_i),
\end{equation}
where $g_{\sigma i}=x_s A_i g_{\sigma N}$ is the $\sigma$-cluster
coupling ($ x_s$ being the $ \sigma $-cluster coupling fraction
calibrated in \cite{Custodio2020}), $ B_i^0 $ is the tabulated vacuum
binding energy of light cluster $ i $  and $ \delta B_i $ is the
binding energy shift, 

which was first defined in Ref.~\cite{Pais2018} as:
\begin{equation}\label{key3}
	\delta B_i=\frac{Z_i}{n_0}(\epsilon_p^*-m n_p^*)+ \frac{N_i}{n_0}(\epsilon_n^*-m n_n^*) \, .
\end{equation}
The total binding energy $ B_i $ is given by
\begin{equation}\label{bcl}
B_i= A_i m^* - M_i^*,\hspace{0.1cm}i={}^{2}\text{H},{}^{3}\text{H},{}^{3}\text{He},{}^{4}\text{He},{}^{6}\text{He}	\, .
\end{equation} 
In the above expressions, $ Z_i $, $ N_i $ are the number of protons and neutrons, respectively, and $ m^*=m-g_{\sigma N}\sigma $ is the nucleon effective mass. The gas energy density $\epsilon_i^*$ and nucleonic density $ n_i^* $, are given by
\begin{eqnarray}
	\epsilon_i^*&=&\frac{1}{\pi^2}\int_0^{p_{F_i}(\rm gas)} p^2 e_i(p) (f_{i+}(p)+f_{i-}(p)) dp \label{enerstar_clusters} \\
	n_i^* &=&\frac{1}{\pi^2}\int_{0}^{p_{F_i}(\rm gas)}  p^2 (f_{i+}(p)+f_{i-}(p)) dp \label{rhostar_clusters} ,
\end{eqnarray}
where $p_{F_i}(\text{gas})=(3 \pi^2 n_i)^{1/3} $ is the Fermi momentum of nucleon $i$ defined using the zero temperature relation between the density and the Fermi momentum, $f_{i\pm}$ are the usual Fermi distribution functions for the particles and anti-particles, and $e_i=\sqrt{p_i^2+m^{*2}}$ is
the corresponding single-particle energy of the nucleon $i$.

\subsection{The light hyperclusters}

The light hyperclusters are introduced in a similar way as the purely nucleonic light clusters. The Lagrangian density for the fermionic light hypercluster $ j= $ ${}^{3}_{\Lambda}\text{H}$ is given by:
\begin{equation}\label{key4}
	\mathcal{L}_{j}=\bar{\Psi}_{j}\left[\gamma_{\mu}i D_{j}^{\mu} - M_{j}^* \right]\Psi_{j} \, , j={}^{3}_{\Lambda}\text{H} \, ,
\end{equation} 
with
\begin{equation}\label{key5}
	i D_{j}^{\mu}=i \partial^{\mu} - g_{\omega j} \omega^{\mu} -  g_{\phi j} \phi^{\mu} \, ,
\end{equation}
whereas for the bosonic light hyperclusters, $ j= $ $^{4}_{\Lambda}\text{H}$, $^{4}_{\Lambda}\text{He}$, the Lagrangian density is given by
\begin{eqnarray}
	\mathcal{L}_j&=& \frac{1}{2}\left( iD_{j}^{\mu} \Psi_{j} \right)^*\left( iD_{\mu j} \Psi_{j} \right)  \\ && - \frac{1}{2}\Psi_{j}^*(M_j^*)^2\Psi_{j} ,\hspace{0.1cm}j={}^{4}_{\Lambda}\text{H} ,{}^{4}_{\Lambda}\text{He}\nonumber \, ,
\end{eqnarray}
with 
\begin{equation}\label{key6}
	i D_{j}^{\mu}=i \partial^{\mu} - g_{\omega j} \omega^{\mu} -  g_{\phi j} \phi^{\mu}- g_{\rho j}\vec{I}_{j}\cdot \vec{\rho^{\mu}} \, .
\end{equation}
The coupling of the hyperclusters to the $\omega$ meson is defined as
\begin{equation}\label{hypercluster_omega_coupling}
	g_{\omega j}=(A_j-1)g_{\omega N} + g_{\omega \Lambda} \, ,
\end{equation}
and the coupling to the $\phi$ meson is the same as the one of the $\Lambda-$hyperon coupling, $ g_{\phi j}=g_{\phi \Lambda} $ (``ideal mixing'' case). The coupling to the $\rho-$meson is the same as the coupling to the nucleons, $g_{\rho j}=g_{\rho N}$.

Following Eq.(\ref{effective_mass_clusters}), the effective mass of the hypercluster $ j $ is given by
\begin{equation}\label{key7}
	M_j^*=  (A_j-1)m + m_{\Lambda} - g_{\sigma j}\sigma -(B_j^0 + \delta B_j) \, 
\end{equation}
with 
\begin{equation}\label{hypercluster_sigma_coupling}
	g_{\sigma j}=x_s( (A_j-1) g_{\sigma N} + g_{\sigma \Lambda}) 
\end{equation}
the $\sigma$-hypercluster coupling. Let $M_j$ be the vacuum mass of a hypercluster $ j $
\begin{equation}\label{mj}
	M_j= (A_j-1)m + m_{\Lambda}- B_j^0  \, .
\end{equation}
The vacuum masses for the hypertriton, $ {}^{3}_{\Lambda}\text{H} $,
the  hyper-hidrogen4, $ {}^{4}_{\Lambda}\text{H} $, and the
hyperhelium4, $ {}^{4}_{\Lambda}\text{He} $, were obtained from
\cite{Adam2019},\cite{Esser2014},\cite{Yamamoto2015}, respectively: $
M_{{}^{3}_{\Lambda}\text{H}}=2990.89$~MeV, $
M_{{}^{4}_{\Lambda}\text{H}}=3922.49$~MeV, $ M_{{}^{4}_{\Lambda}\text{He}}=3921.70$~MeV. From Eq.~(\ref{mj}), the vacuum binding energies can be extracted, $ B_{{}^{3}_{\Lambda}\text{H}}^0=2.793 $ MeV, $ B_{{}^{4}_{\Lambda}\text{H}}^0=10.198 $ MeV, and $ B_{{}^{4}_{\Lambda}\text{He}}^0=10.981$ MeV.

 From \cite{She2020} we see that the experimental production ratio between $ {}^{3}_{\Lambda}\text{H} $ and $ {}^{3}_{}\text{H} $ ($ {}^{3}_{\Lambda}\text{H}/{}^{3}_{}\text{H} $) is $ 0.75\pm0.07 $. Since this ratio must be proportional to the square of the interaction ($(g_{\sigma {}^{3}_{\Lambda}\text{H}}/g_{\sigma {}^{3}_{}\text{H}} )^2 \sim 0.76$), we conclude that the way we define the $\sigma$-hypercluster couplings is in good accordance with experimental data.  

Following the same procedure for the hyperclusters, their binding energy shift is given by
\begin{equation}\label{key8}
	\delta B_j\!=\!\frac{Z_j}{n_0}\!(\epsilon_p^*\!-m n_p^*)+\frac{N_j}{n_0}\!(\epsilon_n^*\!-m n_n^*) + \frac{ \Lambda_j}{n_0}\!(\epsilon_{\Lambda}^*\!-m_{\Lambda}n_{\Lambda}^*) \, ,
\end{equation}
where $\Lambda_j$ is the number of $\Lambda$ hyperons present at each
hypercluster $ j $, which, for the present hyperclusters we consider,
is always equal to 1. The energy density $\epsilon_{\Lambda}^*$ and
the density $n_{\Lambda}^*$ are similar to
Eqs.~(\ref{enerstar_clusters}) and (\ref{rhostar_clusters}),
respectively, and the total binding energy for each hypercluster $j$,
$B_j $ is given by
\begin{equation}\label{bj}
	B_j=  (A_j-1)m^* + m^*_{\Lambda} - M^*_j \, .
\end{equation}

\subsection{The mesonic fields}

The Lagrangian density for the fields have the standard RMF expressions:
\begin{align} 
	\mathcal{L}_{\sigma}&=\frac{1}{2}\partial_{\mu}\sigma\partial^{\mu}\sigma - \frac{1}{2}m_{\sigma}^2 \sigma^2 \, ,\\	
	\mathcal{L}_{\omega}&=- \frac{1}{4}\Omega^{\mu\nu}\Omega_{\mu\nu} + \frac{1}{2}m_{\omega}^2\omega_{\mu}\omega^{\mu} \, ,\\
	\mathcal{L}_{\phi}&=-\frac{1}{4} P^{\mu\nu} P_{\mu\nu}+ \frac{1}{2}m_{\phi}^2\phi_{\mu}\phi^{\mu} \, ,\\
	\mathcal{L}_{\rho}&= - \frac{1}{4}\vec{R}^{\mu\nu} \cdot \vec{R}_{\mu\nu} + \frac{1}{2}m_{\rho}^2\vec{\rho}_{\mu} \cdot \vec{\rho}^{\mu}	\, ,
\end{align}
with $ \Omega_{\mu\nu}=\partial_{\mu} \omega_{\nu}-\partial_{\nu}
\omega_{\mu} $, $ P_{\mu\nu}=\partial_{\mu} \phi_{\nu}-\partial_{\nu}
\phi_{\mu} $, and $ \vec{R}_{\mu\nu}=\partial_{\mu}
\vec{\rho}_{\nu}-\partial_{\nu} \vec{\rho}_{\mu} +g_\rho(\vec{\rho}_\mu \times \vec{\rho}_\nu) $.

We treat the binding energy shifts, $\delta B_{i}$, as in \cite{Typel2009}: we replace the density dependence of these quantities by a vector meson dependence. This is equivalent, in our present study,  to consider in the shifts $\delta B_{i}$ the neutron, proton, and $\Lambda$ densities replaced by
\begin{eqnarray}
	n_n&=& \frac{m_{\omega}^2} {2g_{\omega
                  N}}\omega_0-\frac{g_{\omega\Lambda
                  }}{g_{\phi\Lambda}}
                  \frac{m_{\phi}^2} {2g_{\omega N}} \phi_0
                  - \frac{m_{\rho}^2}{g_{\rho N}}\rho_{03} \, , \\
	n_p&=& \frac{m_{\omega}^2}{2g_{\omega
                  N}}\omega_0-\frac{g_{\omega\Lambda }
                  }{g_{\phi\Lambda}}\frac{m_{\phi}^2}{2g_{\omega N}}
                  \phi_0  + \frac{m_{\rho}^2}{g_{\rho N}} \rho_{03} \, ,  \\
	n_{\Lambda}&=& \frac{m_{\phi}^2 }{g_{\phi\Lambda}} \phi_0 \, .
\end{eqnarray}

With the inclusion of the binding energy shift for each cluster and hypercluster, the equations for the fields read:

\begin{eqnarray}
	m_{\sigma}^2 \sigma &=& \sum_{b}g_{\sigma b}n_b^s + \sum_{i}g_{\sigma i}n_i^s + \sum_{j}g_{\sigma j}n_j^s  \, ,
\end{eqnarray}

\begin{eqnarray}
	m_{\rho}^2 \rho_{03} &=& g_{\rho N}\left[ \sum_{b}I_{3b}n_b + \sum_{i}I_{3i}n_i + \sum_{j}I_{3j}n_j  \right]  \\ 
		 &-&\frac{m_{\rho}^2}{g_{\rho N}n_0}   \!\left(\!  -\frac{\partial\epsilon_n^*}{\partial n_n}\!  + \!\frac{m\partial n_n^*}{\partial n_n}\!  \right) \!\big(n_{{}^{3}\text{H}}^s\!+\!2 n_{{}^{6}\text{He}}^s\!+\!n_{{}^{4}_{\Lambda}\text{H}}^s\big)\! \nonumber \\
	  &-& \frac{m_{\rho}^2}{g_{\rho N}n_0}   \!\left( \frac{\partial\epsilon_p^*}{\partial n_p}\!  - \!\frac{m\partial n_p^*}{\partial n_p} \right) \big(n_{{}^{3}\text{He}}^s\!+\!n_{{}^{4}_{\Lambda}\text{He}}^s\big)\! \nonumber \\
	  &-&\frac{m_{\rho}^2}{g_{\rho N}n_0}
	 \left(\! -\frac{\partial\epsilon_{n}^*}{\partial n_n}\! + \!\frac{m\partial n_{n}^*}{\partial n_n}\! + \!\frac{\partial\epsilon_{p}^*}{\partial n_p}\! - \!\frac{m\partial n_{p}^*}{\partial n_p}\! \right) \nonumber \\ 	 
 &\times& \left(\sum_i n_i^s+\sum_j n_j^s \right) \, , \nonumber	
\end{eqnarray}	

\begin{eqnarray}
	m_{\omega}^2 \omega_0 &=& \sum_{b}g_{\omega b} n_b + \sum_{i}g_{\omega i}n_i + \sum_{j}g_{\omega j}n_j \\ 
	&-& \frac{m_{\omega}^2}{2g_{\omega N}n_0}  \!\left(  \!\frac{\partial\epsilon_n^*}{\partial n_n}\!  - \!\frac{m\partial n_n^*}{\partial n_n}  \right) \left(n_{{}^{3}\text{H}}^s\!+\!2 n_{{}^{6}\text{He}}^s\!+\! n_{{}^{4}_{\Lambda}\text{H}}^s \right) \nonumber \\ 
	&-& \frac{m_{\omega}^2}{2g_{\omega N}n_0} \!\left(  \frac{\partial\epsilon_p^*}{\partial n_p}\!  - \!\frac{m\partial n_p^*}{\partial n_p} \right) \big(n_{{}^{3}\text{He}}^s\!+\! n_{{}^{4}_{\Lambda}\text{He}}^s\big)\!  \nonumber \\
		&-& \frac{m_{\omega}^2}{2g_{\omega N}n_0} \!\left(\frac{\partial\epsilon_{n}^*}{\partial n_n}\! - \!\frac{m\partial n_{n}^*}{\partial n_n}\! + \!\frac{\partial\epsilon_{p}^*}{\partial n_p}\! - \!\frac{m\partial n_{p}^*}{\partial n_p}\! \right)\! \nonumber  \\ 
&\times & \left(\sum_i n_i^s+\sum_j n_j^s \right) \nonumber \, ,
\end{eqnarray}

\begin{eqnarray}
	m_{\phi}^2\phi_0 &=& \sum_{\substack{b=\Lambda,\Sigma^{-,0,+}, \\ \Xi^{-,0}}} g_{\phi b}n_b + \sum_{j} g_{\phi j}n_j \\ 
			&+&\frac{g_{\omega \Lambda}m_{\phi}^2}{2g_{\omega N}g_{\phi\Lambda}n_0} \!\left(  \!\frac{\partial\epsilon_n^*}{\partial n_n}\!  - \!\frac{m\partial n_n^*}{\partial n_n}  \right) \!\big(n_{{}^{3}\text{H}}^s\!+\!2 n_{{}^{6}\text{He}}^s\!+\! n_{{}^{4}_{\Lambda}\text{H}}^s\big)\! \nonumber \\ 
		&+&\frac{g_{\omega \Lambda}m_{\phi}^2}{2g_{\omega N}g_{\phi\Lambda}n_0}  \!\left(   \frac{\partial\epsilon_p^*}{\partial n_p}\!  - \!\frac{m\partial n_p^*}{\partial n_p} \right) \! \big(n_{{}^{3}\text{He}}^s\!+\! n_{{}^{4}_{\Lambda}\text{He}}^s\big)\! \nonumber   \\ 
		&-&\frac{m_{\phi}^2}{g_{\phi\Lambda}n_0} \!\left(\frac{\partial\epsilon_{\Lambda}^*}{\partial n_{\Lambda}}\! - \!\frac{m_{\Lambda}\partial n_{\Lambda}^*}{\partial n_{\Lambda}} \right) \!\big(n_{{}^{3}_{\Lambda}\text{H}}^s\! + \! n_{{}^{4}_{\Lambda}\text{H}}^s\!+ \! n_{{}^{4}_{\Lambda}\text{He}}^s\big)\! \nonumber \\
		&+&\frac{g_{\omega \Lambda}m_{\phi}^2}{2g_{\omega N}g_{\phi\Lambda}n_0} 
	\!\left(\frac{\partial\epsilon_{n}^*}{\partial n_n}\! - \!\frac{m\partial n_{n}^*}{\partial n_n}\! + \!\frac{\partial\epsilon_{p}^*}{\partial n_p}\! - \!\frac{m\partial n_{p}^*}{\partial n_p}\! \right)\! \nonumber  \\ 
&\times& \left(\sum_i n_i^s+\sum_j n_j^s \right) \, , \nonumber   
\end{eqnarray}
where $ I_{3b} $, $ I_{3i} $, $ I_{3j} $ correspond to the isospin projections of the baryons $ b $, light clusters $ i $ and light hyperclusters $ j $, respectively. The quantities $n_b$, $ n_i $, $ n_j $ correspond to the particle's densities, whereas  $n_b^s$, $ n_i^s $, $ n_j^s $ represent their scalar densities.

\subsection{Chemical Equilibrium}

In our system, the charge fraction $ Y_Q $ is fixed and defined as:
\begin{equation}\label{key9}
	Y_Q=\sum_{b}q_bY_b+\sum_{i}\frac{q_i}{A_i}Y_i+\sum_{j}\frac{q_j}{A_j}Y_j 
\end{equation} 
where $ q_b $, $ q_i $, $ q_j $ are the electric charges of baryon $ b $, light cluster $ i $ and light hypercluster $ j $, respectively. The quantities $ Y_b $, $ Y_i $ and $ Y_j $ correspond to the mass fractions of the different particles and are given by:
\begin{equation}\label{key10}
	Y_b=\frac{n_b}{n_B},\hspace{0.1cm}Y_i=A_i\frac{n_i}{n_B},\hspace{0.1cm}Y_j=A_j\frac{n_j}{n_B},
\end{equation}
where $ n_B $ is the total density of the system.

The chemical potential $\mu_b$ of baryon $ b $ can be written as:
\begin{equation}\label{key11}
	\mu_b=\mu_n-q_b\mu_e
\end{equation}
where $\mu_n$, $\mu_e$ are the neutron and electrical charge chemical potentials, respectively. Since $\mu_e\!=\!\mu_n\!-\!\mu_p$, the hyperon chemical potentials can be written in terms of the nucleons chemical potentials: $\mu_{\Lambda}\!=\!\mu_n$, $\mu_{\Sigma^-}\!=\!2\mu_n - \mu_p$, $\mu_{\Sigma^0}\!=\!\mu_n$, $\mu_{\Sigma^+}\!= \!\mu_p$, $\mu_{\Xi^-}\!=\!2\mu_n\! - \!\mu_p$, $\mu_{\Sigma^0}\!=\!\mu_n$.

For a light cluster $ i $, their chemical potential $\mu_i$ can also be defined as a function of $\mu_n$ and $\mu_p$:
\begin{equation}\label{key12}
	\mu_{i}=N_i\mu_n + Z_i\mu_p
\end{equation}
whereas for a light hypercluster $ j $, $\mu_{\Lambda}$ also needs to be taken into account:
\begin{equation}\label{key}
	\mu_{j}=N_j\mu_n + Z_j\mu_p + \Lambda_j \mu_{\Lambda}.
\end{equation}

The effective chemical potential $\mu_c^*$ of any particle $ c=b,i,j $ present in our system can be written in terms of its chemical potential $\mu_c$ as:
\begin{equation}\label{key13}
\mu_c^*= \mu_c - g_{\omega c}\omega_0 - g_{\phi c}\phi_0-g_{\rho c}I_{3c}\rho_{03} - A_c\Sigma_0^R	
\end{equation}
where $ \Sigma_0^R $ is the rearrangement term present in models with density-dependent couplings in order to guarantee thermodynamical consistency:
\begin{eqnarray}
\Sigma_0^R\!=\!\! \sum_{c}\Big(&&\frac{\partial g_{\omega c}}{\partial n_B}\omega_0 n_c\! + \!\frac{\partial g_{\phi c}}{\partial n_B}\phi_0 n_c\! + \!I_{3c}\frac{\partial g_{\rho c}}{\partial n_B}\rho_{03} n_c  \nonumber \\ &&-  \frac{\partial g_{\sigma c}}{\partial n_B}\sigma_0 n_c^s \Big). 
\end{eqnarray}

\begin{figure*}[!t]
	\includegraphics[width=0.75\linewidth]{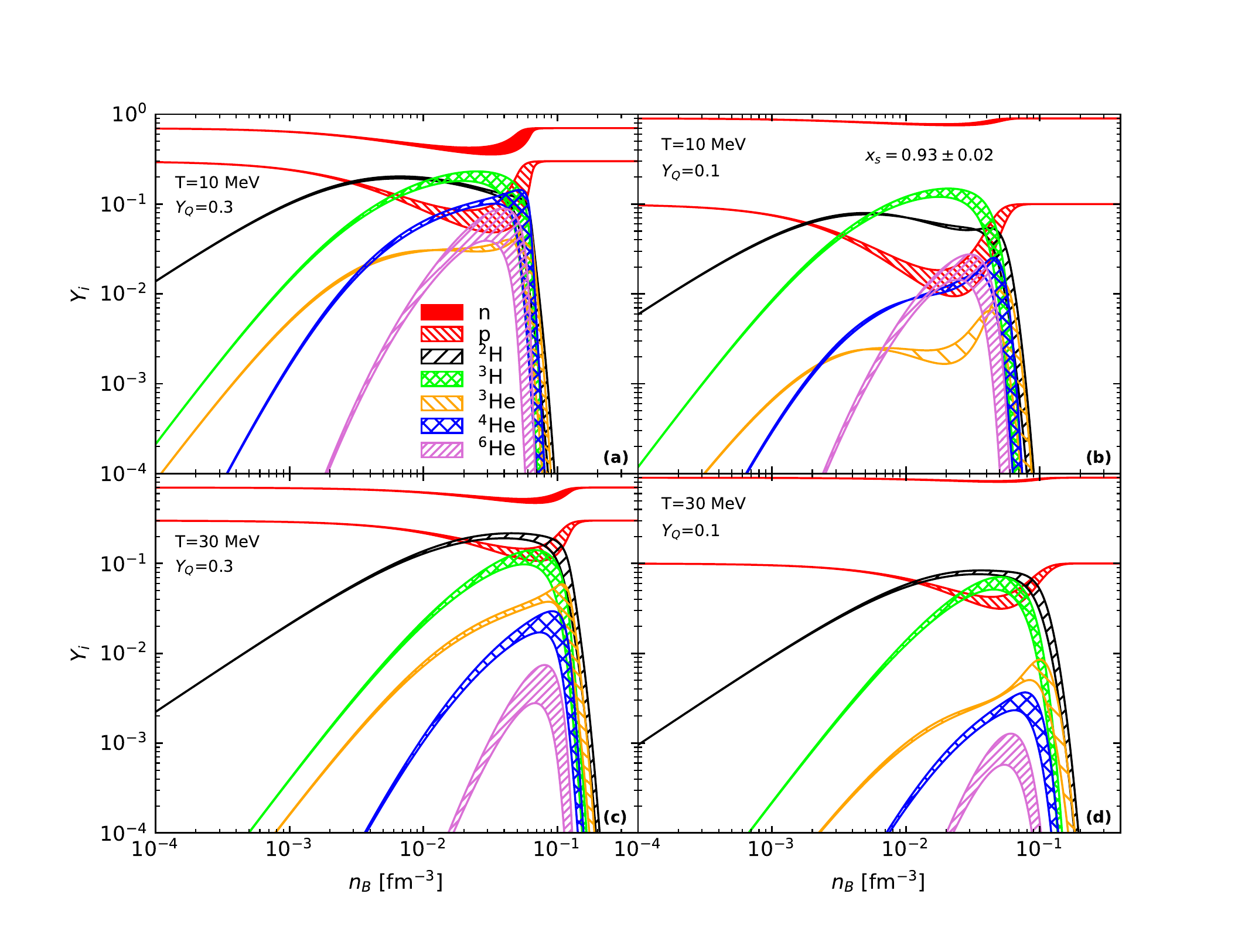} 
	\caption{Mass fractions of light clusters  ($^2$H, $^3$H, $^3$He, $^4$He
		and $^6$He) and unbound protons and
		neutrons  in equilibrium are plotted  versus the  density for $T=10$ MeV (top) and 30 MeV
		(bottom) with charge fraction of $Y_Q=0.3$ (left) and 0.1 (right). The
		bands take into account the uncertainty on the $x_s$ coupling fraction of the
		clusters to the $\sigma$-meson.}
	\label{fig1}
\end{figure*}

\section{Results\label{sec3}}

\begin{figure*}[!t]
	\includegraphics[width=.9\linewidth]{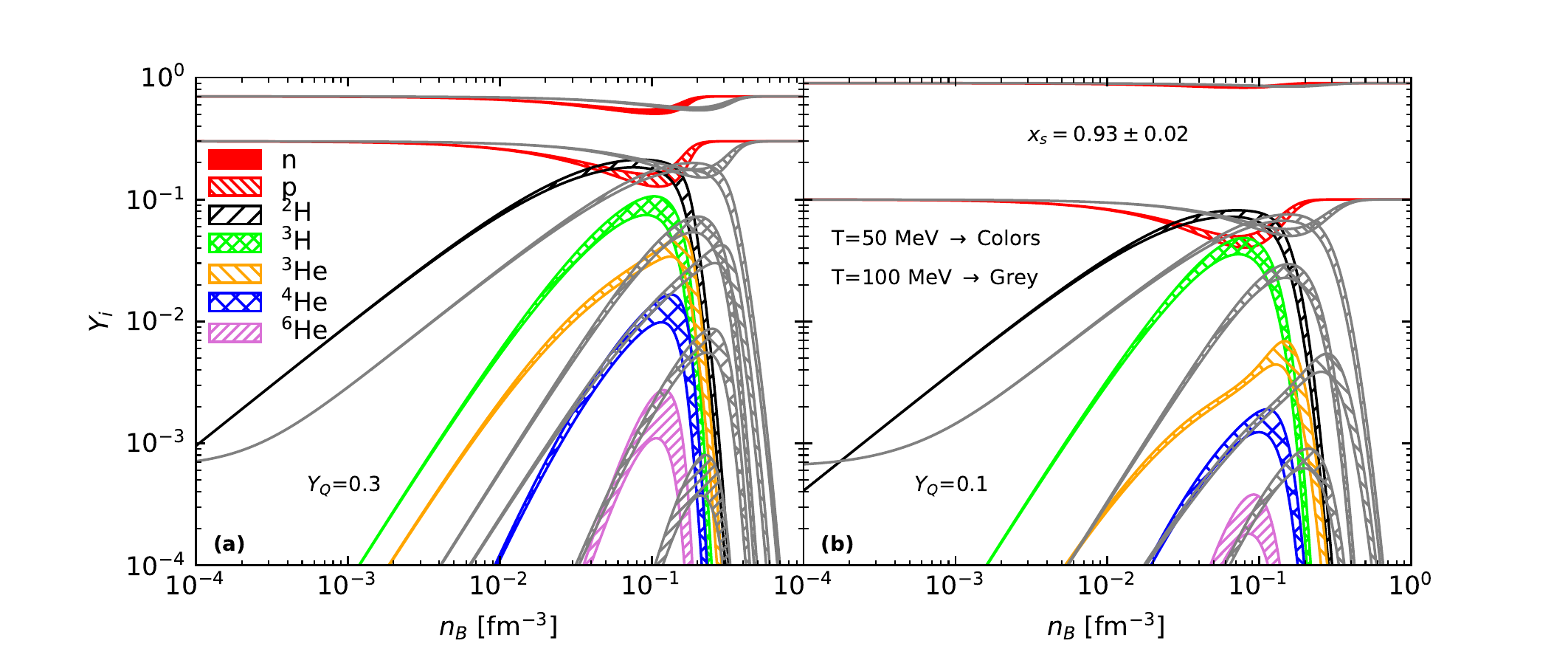} 
	\caption{Mass fractions of light clusters  ($^2$H, $^3$H, $^3$He, $^4$He
		and $^6$He) and unbound protons and
		neutrons  in equilibrium are plotted  versus the  density for $T=50$
		 MeV (colored lines) and 100 MeV
		(grey lines), with a charge fraction of $Y_Q=0.3$ (left) and 0.1 (right). The
		bands take into account the uncertainty on the $x_s$ coupling fraction of the
		clusters to the $\sigma$-meson. }
	\label{fig2}
\end{figure*}

In the present section we discuss how the presence of light clusters
affects the abundances of heavy baryons at low densities and
temperatures $T\lesssim 50$ MeV, and two different charge
fractions, $ Y_Q=0.1$ and 0.3. Above  the critical temperature, $T_c\approx 15$ MeV, we
do not expect the presence of heavy clusters, so, and as mentioned in the previous Sections, we consider 5 light clusters, $^2$H, $^3$H, $^3$He, $^4$He and $^6$He, which were measured by INDRA
\cite{indra}, and three light hypernuclei $^3_\Lambda$H, $^4_\Lambda$H,
$^4_\Lambda$He. All the calculations shown are for the DD2 RMF model \cite{Typel2009}. In Ref.~\cite{Custodio2020}, the cluster-meson $\sigma$ coupling fraction $x_s$ was  calibrated
to the equilibrium constants obtained in \cite{indra} for different RMF models. For the
density-dependent DD2 RMF model, a value of $x_s=0.93\pm 0.02$ was obtained. This range of values is going to be used throughout this work.

\begin{figure}[!t]
	\includegraphics[width=1\linewidth]{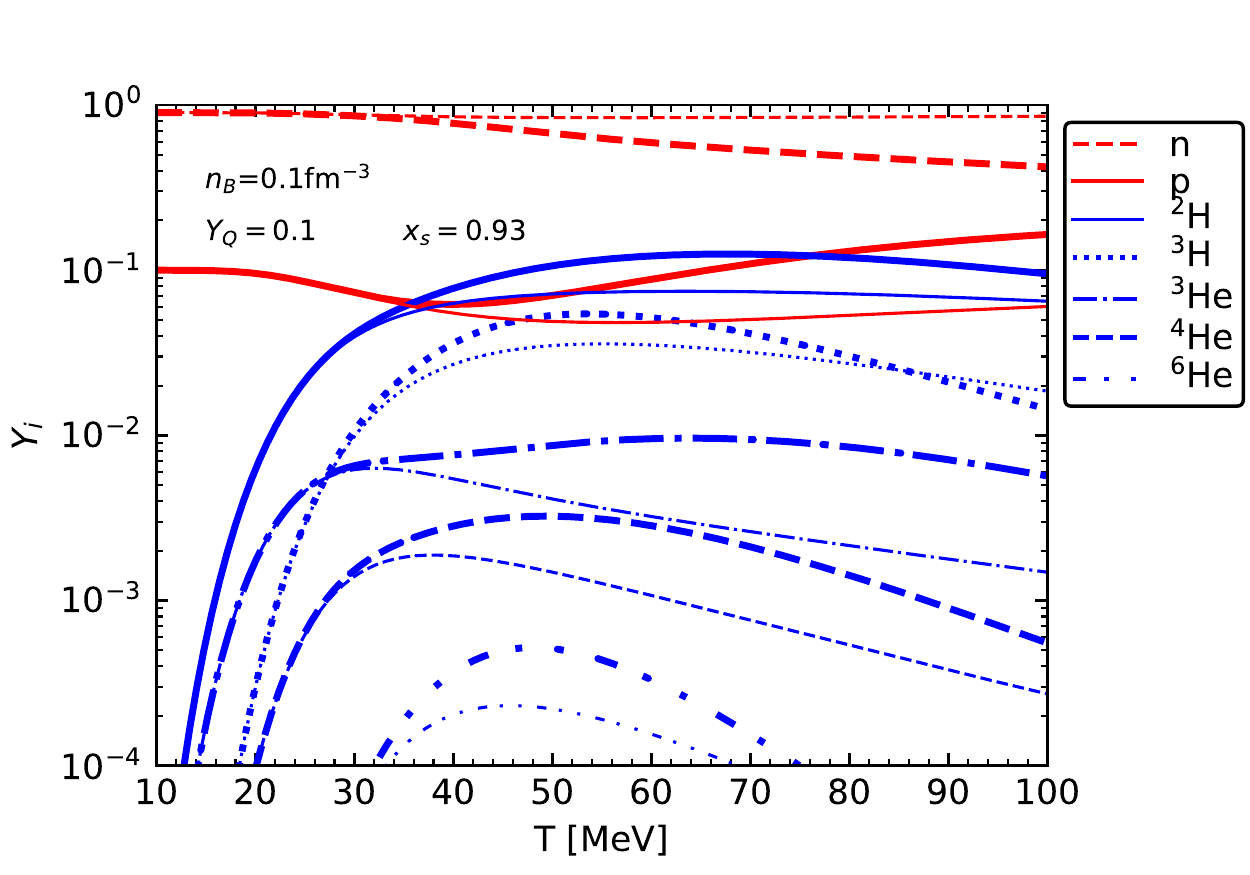}
	\caption{Unbound nucleon and light cluster fractions in a calculation 
		with (thick lines) and without (thin lines)  hyperons as a function of the temperature for a charge fraction of
		$Y_Q=0.1$, and a density of $n_B=0.1$ fm$^{-3}$. The scalar cluster-meson coupling is fixed to $x_s=0.93$.}
	\label{fig2a}
\end{figure}

In Fig. \ref{fig1}, we plot the mass fractions of light clusters  ($^2$H, $^3$H, $^3$He, $^4$He
and $^6$He) and unbound protons and
neutrons  in equilibrium as a function of  density for two temperatures $T=10$ MeV (top) and 30 MeV
(bottom) and two different values of the  charge fraction $Y_Q=0.3$
(left) and 0.1 (right).  The bands take into account the uncertainty on  $x_s$, and mainly affect 
the fraction maximum and the dissolution density. Several comments are in order concerning the effect of the temperature and charge: i) at the lowest densities, it is the mass that
determines the most abundant light cluster, and the smaller the mass
the larger the abundance; ii) for $T=10$ MeV, the most abundant cluster
at the fraction maximum is the tritium, reflecting the isospin
asymmetry. In particular, $^6$He becomes
more abundant than $^3$He for the two charge fractions considered; iii) at $T=30$ MeV, the mass defines the largest abundances; iv) for $T=10$ MeV, it is interesting to observe that   even though
$^3$He is less abundant than $^3$H, $^4$He or $^6$He,  it dissolves at larger
densities. This is an effect of the binding energy shift that depends
on the density of unbound neutrons and protons separately. The
neutrons, being more abundant, have a stronger effect, and, in
particular, affect more the clusters with a larger neutron fraction; v) for large temperatures, here represented by $T=30$ MeV,
the deuteron is the most abundant for all densities due to its smaller
mass. Moreover, at the maximum of the cluster fractions, their mass
fractions are larger than the proton fraction.

\begin{figure*}[!t]
	\includegraphics[width=0.9\linewidth]{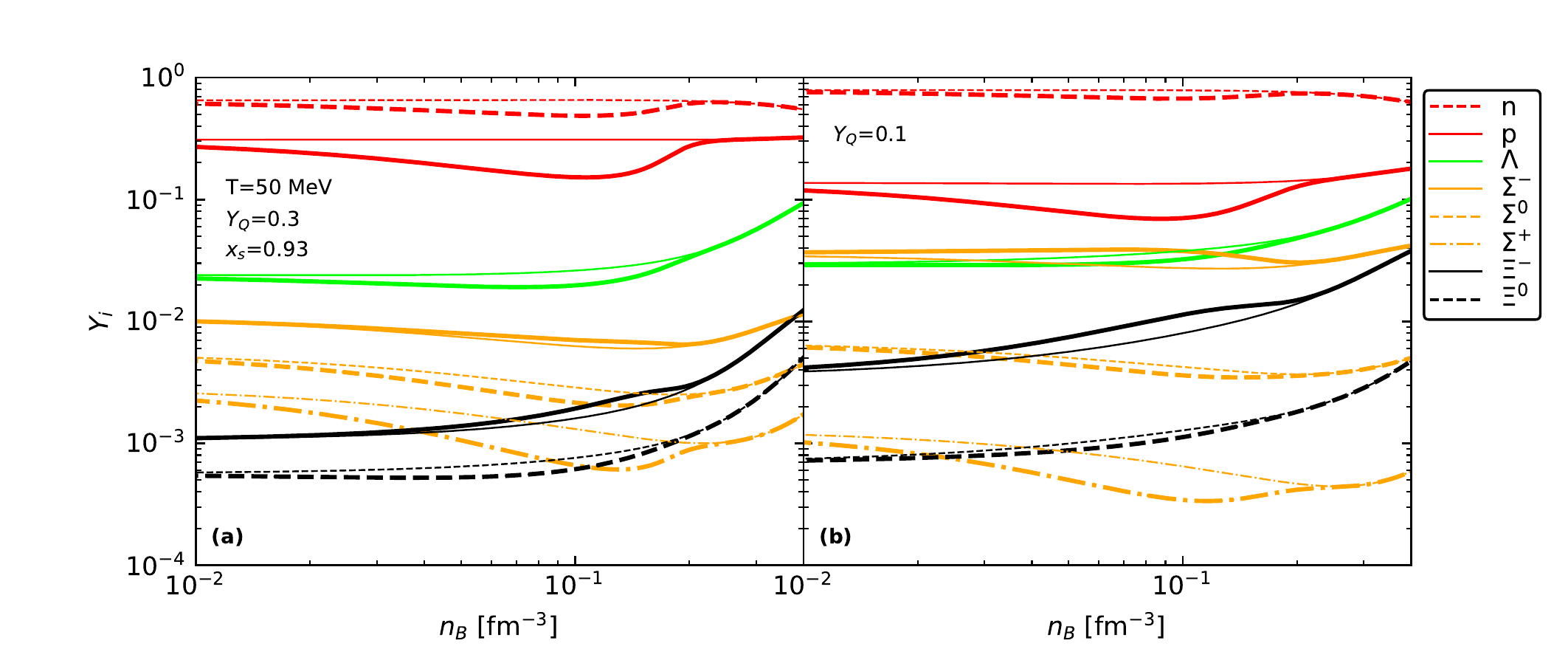} 
	\caption{Unbound nucleon and hyperon fractions as a function of the density in a calculation 
		with (thick lines) and without (thin lines)  light
		clusters, for a charge fraction of $Y_Q=0.3$ (left) and 0.1 (right) and $T=50$ MeV. The scalar cluster-meson coupling fraction is set to $x_s=0.93$.}
	\label{fig3}
\end{figure*}

In Fig. \ref{fig2}, the mass fractions are again plotted against the density, but this time for two
larger temperatures, $T=50$ and 100 MeV, the last one represented by
grey lines to be well distinguished from the $T=50$ MeV case.  As already discussed for
$T=30$ MeV, the relative abundances of the light nuclei are dictated
by their masses, the deuteron being the most abundant and $^6$He the least. The superposition of the distribution for both temperatures shows clearly that an increase of the temperature pushes the
light-nuclei maxima to larger densities and reduces the abundances of
the heavier clusters: only the deuteron keeps a similar fraction at
the maximum. The cluster dissolution shifts to much larger densities
for the larger temperature. A reduction of the charge fraction reduces
the cluster fractions: for $T=100$ MeV, $ Y_Q=0.1 $ the fraction of $^6$He is
always below 10$^{-4}$.

 In order to better understand the effect of the
temperature, we show in Fig. \ref{fig2a} the unbound nucleon and cluster abundances as a
function of the temperature for a charge fraction  $Y_Q=0.1$ and a density $n_B=0.1$ fm$^{-3}$. This density value was chosen because it is where the fraction of the clusters is close to a maximum in the range of temperatures considered. It is seen that the abundance of deuterons
surpasses the one of protons for $25\lesssim T\lesssim 70$ MeV. It is also
above $T=25$ MeV that the cluster  fractions obtained with and
without hyperons start differing, and they start being more abundant in the
presence of hyperons.

The effect of the inclusion of light clusters on the hyperon
fractions is clearly seen in Fig. \ref{fig3}:  the thick lines were
obtained including clusters, while the calculation without clusters is
represented by thin lines. The main effect of introducing clusters
is a reduction of the unbound nucleons and of the electrically neutral or
positive hyperons, while the fraction of the negatively charged hyperons
increases. The formation of clusters is energetically favored but
these clusters are positively charged, so its formation is compensated
by a reduction of the unbound nucleons, together with  a reduction (increase)  of
positively (negatively) charged baryons. A decrease of the
neutron fraction also induces a reduction of the other neutral
baryons. Moreover, a smaller charge fraction favors the formation of
negatively charged baryons, and for $Y_Q=0.1$, it is seen a clear
competition between $\Sigma^-$ and $\Lambda$ for the smaller
densities. At smaller densities, for a fixed temperature, the hyperon
mass defines the abundance, but for larger densities, the magnitude and
signal of the hyperon potential is reflected on the hyperon abundances. In
particular, the fraction of $\Xi^-$s which feels an attractive
potential becomes larger than the one of $\Sigma^-$ which feels a repulsive interaction.

\begin{figure*}[!t]
	\includegraphics[width=0.9\linewidth]{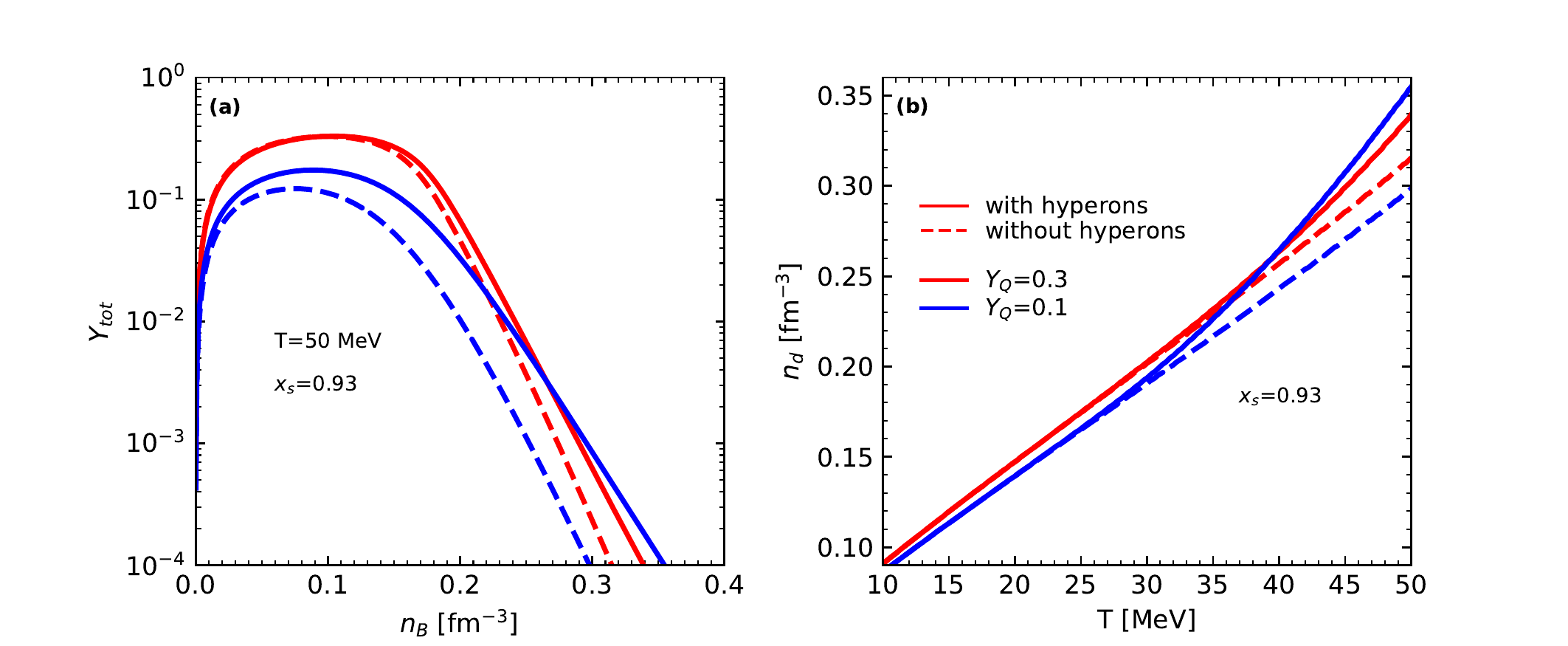} 
	\caption{Total mass fraction of the light clusters  as a function of the density at $T=50$ MeV (left) and the dissolution density of the clusters, $n_d$, as a function of the temperature (right) for a calculation with (solid) and without (dashed) hyperons and a charge fraction of $Y_Q=0.3$ (red) and 0.1 (blue). The scalar cluster-meson coupling fraction is set to $x_s=0.93$. } 
	\label{fig4}
\end{figure*}

The effect of the hyperons on the cluster abundances, which was already seen in Fig.~\ref{fig2a}, and on the dissolution densities is clearly seen in Fig.~\ref{fig4}. In the left panel, we show the total mass fraction of all the light clusters at $T=50$ MeV (notice the linear scale on
the x-axis contrary to the log-scale used in the previous figures), and in the right panel, the dissolution density of the clusters, $n_d$, which was defined as the density for which the cluster
fraction has dropped to 10$^{-4}$ is displayed. The charge fraction is set to $Y_Q=0.3$ and 0.1, and the scalar cluster-meson coupling fraction to $x_s=0.93$. Two different calculations are compared: a calculation with the full baryonic octet (solid lines), and excluding hyperons (dashed lines). The main effects of including hyperons are: i) to increase the cluster fraction above the maximum of
the cluster distribution, shifting the dissolution density to larger
densities, the larger the temperature the stronger the effect; ii) the increase of the dissolution density starts to be non-negligible for $T\gtrsim 25-30$ MeV; iii) the smaller the charge
fraction, the stronger the effect. For $T=50$ MeV, the main effect is an
increase of the dissolution density 
of the order of 10\% if $Y_Q=0.3$, and 20\% for  $Y_Q=0.1$. Since the presence of the hyperons reduces the nucleon fraction, this is reflected on the medium effects felt by the clusters through the binding energy shift that is
smaller. Moreover,  
the couplings to mesons become smaller since the couplings
of the hyperons to the mesons are weaker. This explains why the effect of the hyperons on the clusters is larger for $ Y_Q=0.1 $, since, as we saw in Fig.~\ref{fig3}, a smaller charge fraction corresponds to an overall larger hyperon fraction.

\begin{figure*}[!t]
	\includegraphics[width=1\linewidth]{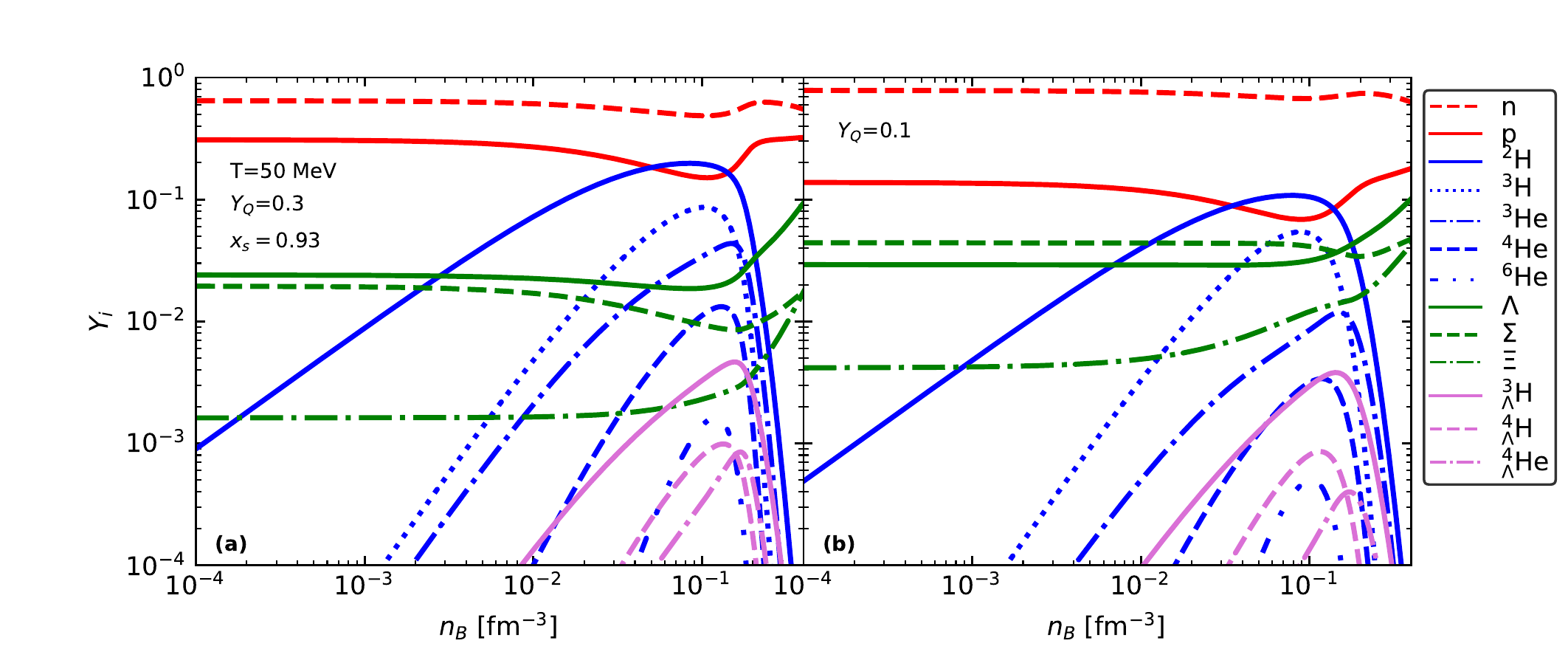} 
	\caption{Mass fractions of the unbound protons and neutrons (red lines), $\Lambda$, $\Sigma$ and $\Xi$ (green lines), light clusters (blue lines) and
		light hypernuclei (pink lines) as a function of the density for $T=50$ MeV and  $x_s=0.93$, with  $Y_Q=0.3$ (left) and 0.1 (right).}
	\label{fig5}
\end{figure*}

\begin{figure*}[!t]
	\includegraphics[width=1\linewidth]{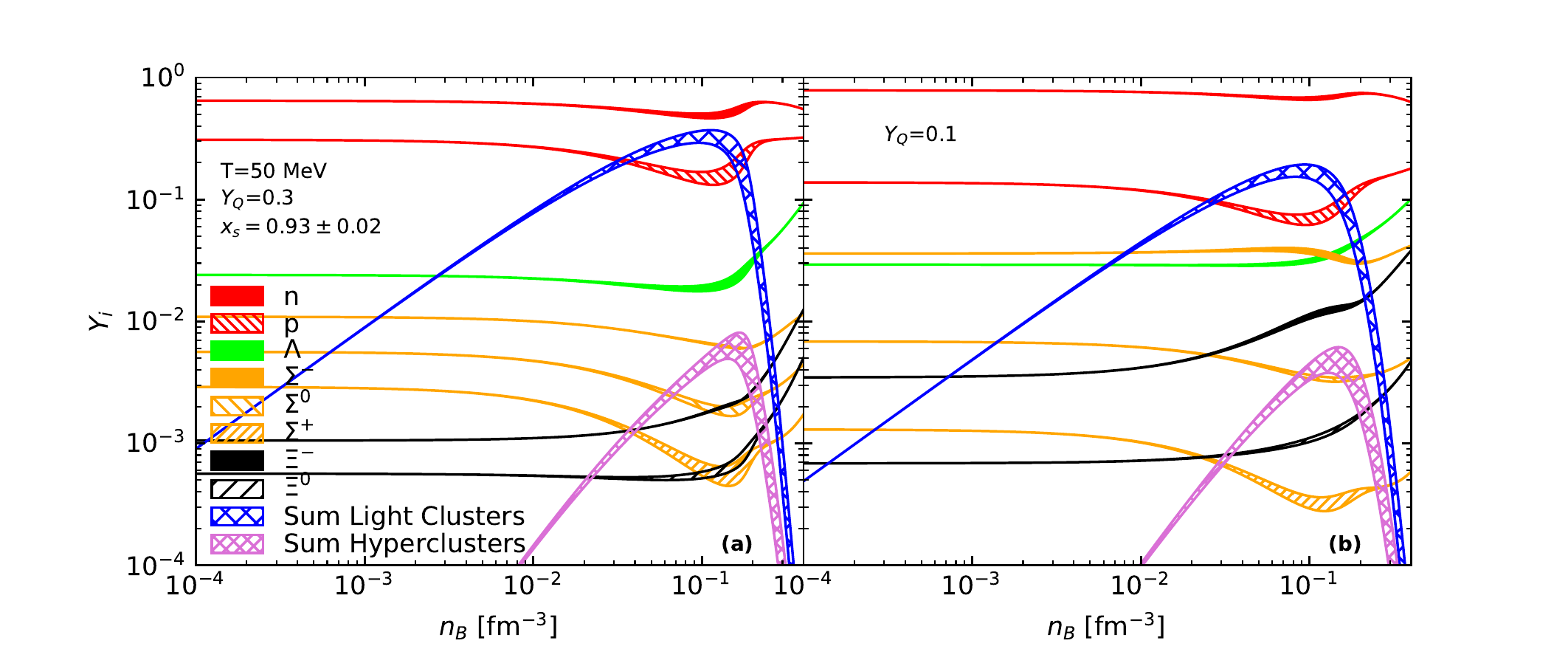} 
	\caption{Mass fractions of the unbound protons and neutrons (red), $\Lambda$ (green), $\Sigma^{-,0,+}$ (orange) and $\Xi^{-,0}$ (black), total light clusters (blue) and
		light hypernuclei (pink) as a function of the density for $T=50$ MeV and  $x_s=0.93 \pm 0.02$, with  $Y_Q=0.3$ (left) and 0.1 (right).}
	\label{fig6}
\end{figure*}

In the following figures, we are going to study the effect of considering hyperclusters in our calculations. As mentioned in the previous sections, we take 
$^3_\Lambda$H , known as hypertriton, $^4_\Lambda$H
(hyperhidrogen 4), and $^4_\Lambda$He (hyperhelium 4).

A fraction of hypernuclei  above 10$^{-4}$ is only obtained
for big enough temperatures, i.e. $T\gtrsim25$ MeV; for lower temperatures,  the abundance of $\Lambda$ hyperons is still too small to give rise to significant hypercluster fractions. Therefore, in the next two Figures, we consider $T=50$ MeV. In Fig. \ref{fig5}, the light nuclei and
hypernuclei mass fractions are plotted together with the unbound
proton and neutron fractions, the $\Lambda$ fraction, the total
$\Sigma$ fraction corresponding to the sum of the $\Sigma^{+,0,-}$
fractions, and the total $\Xi$   fraction corresponding to the sum of the $\Xi^{0,-}$
fractions, for a charge fraction of $Y_Q=0.3$ (left) and 0.1
(right). There is a clear competition between the hypernuclei and the $^4$He and $^6$He light clusters, i.e. the light clusters with a larger mass: for $Y_Q=0.1$, the hypernuclei have larger
abundances,  but even for the larger charge fraction, the dissolution
density occurs at larger densities for the hypernuclei. The behavior
of the hyperclusters in the medium is defined by their couplings to the
mesons. The difference in relation to the light clusters may be attributed to the fact that hypernuclei are interacting more weakly  with the medium, which is clearly seen considering the hypercluster couplings defined in Eqs.~(\ref{hypercluster_omega_coupling}) and (\ref{hypercluster_sigma_coupling}). Since the coupling of the hyperclusters to the $\omega$-meson  is strongly correlated with the dissolution density, a smaller $\omega$-coupling implies larger dissolution densities. On the other
hand, a weaker coupling to the $\sigma$-meson gives rise to smaller
mass fractions, since a smaller binding occurs. Also, the binding energy shift
is weaker for the hypernuclei: this binding shift is introduced to
take into account Pauli blocking, but hypernuclei have less nucleons
and therefore experience smaller shifts.

\begin{figure*}[!t]
	\includegraphics[width=1\linewidth]{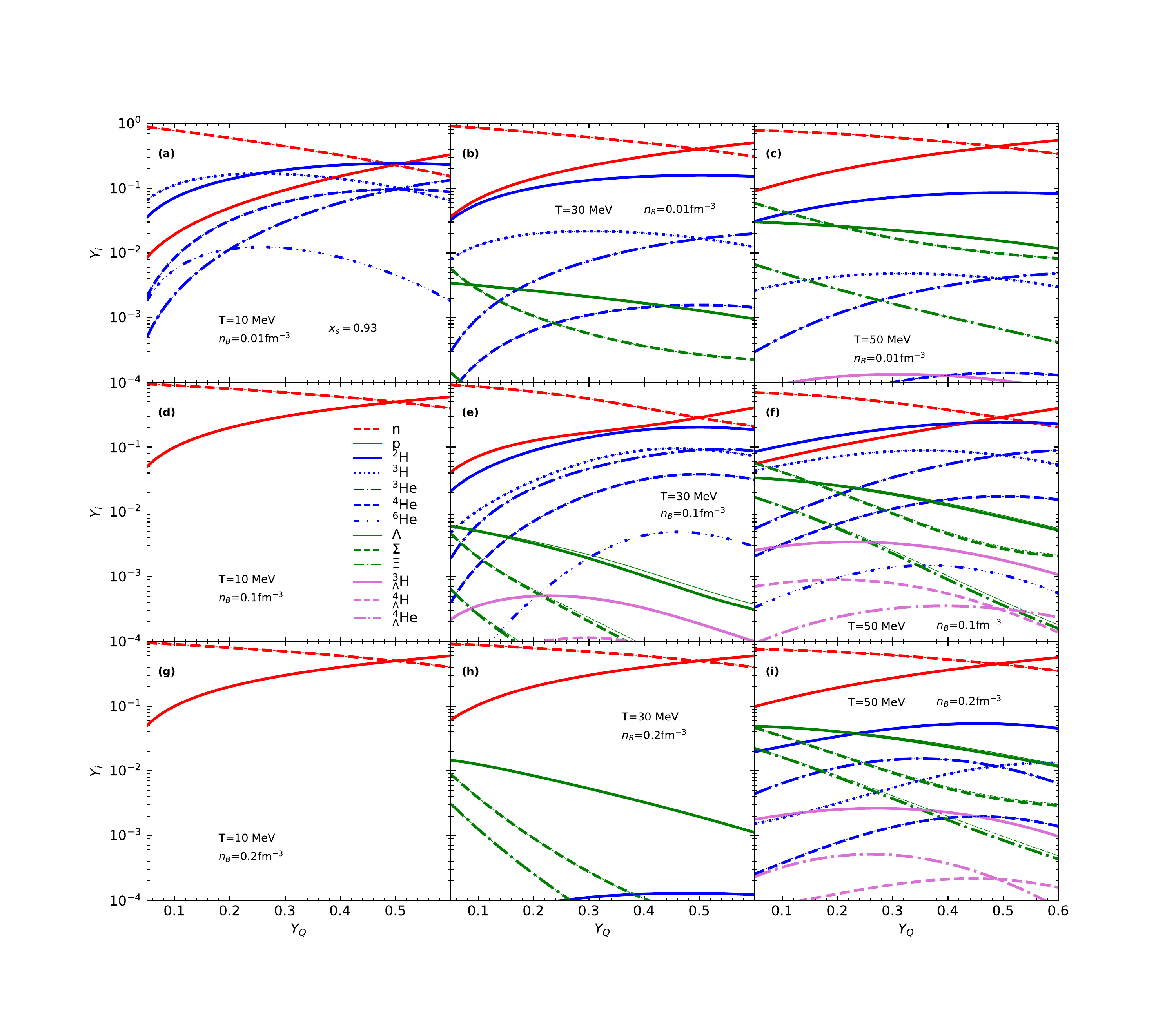} 
	\caption{Mass fractions of the unbound protons and neutrons (red),  unbound
		hyperons: $\Lambda$ (solid green), sum of $\Sigma^{+,0,-}$ (dashed green) and sum of  $\Xi^{-,0}$ (dash-dotted green), light clusters (blue), and light hypernuclei  (pink) as a function of the
		charge fraction for $T=10$ MeV (left), $T=30$ MeV (middle) and $T=50$ MeV (right). The
		fractions were determined at $n_B=0.01$ fm$^{-3}$(top), 0.1 fm$^{-3}$ (middle) and 0.2 fm$^{-3}$(bottom). The scalar cluster-meson coupling is set to $x_s=0.93$.  }
	\label{fig7}
\end{figure*}

It is also interesting to notice that the isospin pair formed by the hyperclusters $^4_\Lambda$H and $^4_\Lambda$He behaves in a similar way to the analogous isospin pair formed by the purely nucleonic clusters $^3$H and $^3$He. In fact, since the $\Lambda$-hyperon present at the hyperclusters has isospin zero, the interactions of these two pairs of clusters with the medium is similar, the only difference being their masses and binding energies, resulting in smaller fractions for the hyperclusters.

In Fig.~\ref{fig6}, the total light cluster fraction and the total light
hypercluster fraction are compared with the baryonic octet fractions
for the two charge fractions, 0.3 and
0.1. We take $T=50$ MeV, and we calculate the effect of the uncertainty on
the $x_s$ coupling of the particle fractions, shown by the bands. The abundances of the
hypernuclei are small compared to the light nuclei, and even taking $T=100$ MeV (not shown), there is not a big
difference whether the hypernuclei are included or not in the
calculation, only slightly affecting the abundances of the heavier
clusters and their dissolution density. 

In order to understand how the charge fraction affects the light
cluster abundances, and under which conditions the hyperclusters are
more abundant, in Fig.~\ref{fig7}, we plot for a fixed density (0.01, 0.1 and
0.2 fm$^{-3}$), and temperatures 10,  30 and 50 MeV, the cluster fractions
as a function of the charge fraction. The densities chosen are below,
close and above the cluster fraction maxima. Depending on the temperature, the last
two density values may be above the dissolution density, taken as the density for which the cluster fraction is below 10$^{-4}$. Considering the lowest density, we conclude that: i) for the
lowest temperature, the most abundant clusters are not only determined by
their mass, but also by their isospin and binding energy, contrary to the other
two temperatures, for which the mass essentially determines their abundances,
and only in a second order, the isospin; ii) hyperons are only present
at $T=30$ and 50 MeV, and hypernuclei appear with an
abundance above 10$^{-4}$ at $T=30$ MeV for $n_B=0.1$~fm$^{-3}$ and at  $T=50$
MeV for all densities considered; iii) protons may be less abundant than
some light clusters, as for instance $^2$H and $^3$H, below $Y_Q<0.5$
for $T=10$ MeV, and $T=50$ MeV and $n_B=0.1$~fm$^{-3}$; 
iv) only for $T=50$ MeV do hyperons become more abundant than most of the
light clusters (only $^2$H  are more abundant).
For $T=10$ MeV and the lowest density, only nucleons
are present, clusters have already dissolved and hyperons did not set
in. For $T=30$ MeV, at $n_B=0.2$ fm$^{-3}$, only deuterons did not
dissolve and  hypernuclei are only  present for 0.1
fm$^{-3}$. For $T=50$ MeV, hypernuclei are present in the three
densities considered although only quite a few for the lowest density. Once again, a similar behaviour is observed for the pairs $^4_\Lambda$H, $^4_\Lambda$He and $^3$H, $^3$He.
Hypernuclei seem to be most abundant for charge fractions of
the order of $Y_Q=0.3$ in all situations studied.

\section{Conclusions\label{sec4}}

The effect of hyperonic degrees of freedom on the low-density EoS of
hot matter, as may occur in events connected with neutron stars, was
studied within the density-dependent DD2 RMF model. 
The study was performed at a fixed charge fraction and considered temperatures until 100 MeV.  The 
degrees of freedom included in the calculations were nucleons, hyperons, and light nuclei and
hypernuclei. The introduction of light clusters was done following the
formalism first described in Ref.~\cite{Pais2018}. Light clusters couple to the
mesonic fields and a binding energy shift is included in order to account for
the Pauli blocking. This contribution essentially influences the
dissolution density of the cluster.  A similar formalism has been
presented and applied in \cite{Typel2009,Hempel2015,Fischer2020}, the difference
being the model description of the coupling of light clusters to the
mesonic degrees of freedom. Moreover, in Ref.~\cite{Typel2009}, the cluster
binding shifts are temperature dependent, with the shifts determined
from a quantum statistical calculation \cite{Ropke2015}.

At low temperatures, the abundances are determined by the cluster
binding energy and isospin, and for charge fractions below 0.3,
light clusters like $^6$He are more abundant than $^3$He or even
$\alpha$-particles. However, neutron-rich clusters dissolve at lower
densities due to the stronger binding energy shifts, which take into account Pauli
blocking effects. Larger temperatures shift the cluster fraction maxima and dissolution densities to larger densities,  they decrease their abundances, except for
the deuteron, and they define the cluster abundances in terms of their
masses, with the light clusters being more abundant.

In this work, we also showed that the presence of hyperons shifts the dissolution of
clusters to larger densities and increases the  cluster abundances for
temperatures $T\gtrsim 25$ MeV. This effect is larger the smaller the
charge fraction, and the higher the temperature. The increase of
clusters is attributed to a weaker effect of the Pauli-blocking
implemented in the model via the binding energy shifts since the overall
nucleon densities is lower.  Besides, the clusters also affect the
hyperon fractions: while neutral and positively charged baryons
decrease when clusters are included, the fraction of negatively
charged hyperons increase.
Hypernuclei set in at
temperatures above 25 MeV, and for $T\gtrsim 50$ MeV, they compete with
$\alpha$-particles and $^6$He. However, switching off the hypernuclei
does not influence much the other particles. It was shown that the
larger abundances for the total fraction of hyperclusters occurs for a
charge fraction close to 0.3.
One expects that a reduction of unbound nuclei and neutral or
positively charged hyperons, and an increase of light clusters, will
affect the  reaction rates that determine the core-collapse supernova
evolution or the binary merger.

In our model, clusters survive up to quite large densities if the
temperatures are high. This must be further investigated and it may be
necessary to include a temperature dependence on the binding energy
shifts. Although one would expect that clusters would dissolve, it is
also true that light clusters survive up to temperatures as high as 150
 MeV  as discussed in Ref.~\cite{Braun-Munzinger2018}.

\section*{ACKNOWLEDGMENTS}
This work was partly supported by the FCT (Portugal) Projects No. UID/FIS/04564/2020 and POCI-01-0145-FEDER-029912, and by PHAROS COST Action CA16214. H.P. acknowledges the grant CEECIND/03092/2017 (FCT, Portugal).

%



\begin{thebibliography}{38}%
\makeatletter
\providecommand \@ifxundefined [1]{%
 \@ifx{#1\undefined}
}%
\providecommand \@ifnum [1]{%
 \ifnum #1\expandafter \@firstoftwo
 \else \expandafter \@secondoftwo
 \fi
}%
\providecommand \@ifx [1]{%
 \ifx #1\expandafter \@firstoftwo
 \else \expandafter \@secondoftwo
 \fi
}%
\providecommand \natexlab [1]{#1}%
\providecommand \enquote  [1]{``#1''}%
\providecommand \bibnamefont  [1]{#1}%
\providecommand \bibfnamefont [1]{#1}%
\providecommand \citenamefont [1]{#1}%
\providecommand \href@noop [0]{\@secondoftwo}%
\providecommand \href [0]{\begingroup \@sanitize@url \@href}%
\providecommand \@href[1]{\@@startlink{#1}\@@href}%
\providecommand \@@href[1]{\endgroup#1\@@endlink}%
\providecommand \@sanitize@url [0]{\catcode `\\12\catcode `\$12\catcode
  `\&12\catcode `\#12\catcode `\^12\catcode `\_12\catcode `\%12\relax}%
\providecommand \@@startlink[1]{}%
\providecommand \@@endlink[0]{}%
\providecommand \url  [0]{\begingroup\@sanitize@url \@url }%
\providecommand \@url [1]{\endgroup\@href {#1}{\urlprefix }}%
\providecommand \urlprefix  [0]{URL }%
\providecommand \Eprint [0]{\href }%
\providecommand \doibase [0]{http://dx.doi.org/}%
\providecommand \selectlanguage [0]{\@gobble}%
\providecommand \bibinfo  [0]{\@secondoftwo}%
\providecommand \bibfield  [0]{\@secondoftwo}%
\providecommand \translation [1]{[#1]}%
\providecommand \BibitemOpen [0]{}%
\providecommand \bibitemStop [0]{}%
\providecommand \bibitemNoStop [0]{.\EOS\space}%
\providecommand \EOS [0]{\spacefactor3000\relax}%
\providecommand \BibitemShut  [1]{\csname bibitem#1\endcsname}%
\let\auto@bib@innerbib\@empty
\bibitem [{\citenamefont {Adam}\ \emph {et~al.}(2020)\citenamefont {Adam} \emph
  {et~al.}}]{Adam2019}%
  \BibitemOpen
  \bibfield  {author} {\bibinfo {author} {\bibfnamefont {J.}~\bibnamefont
  {Adam}} \emph {et~al.} (\bibinfo {collaboration} {STAR}),\ }\href {\doibase
  10.1038/s41567-020-0799-7} {\bibfield  {journal} {\bibinfo  {journal} {Nature
  Phys.}\ }\textbf {\bibinfo {volume} {16}},\ \bibinfo {pages} {409} (\bibinfo
  {year} {2020})},\ \Eprint {http://arxiv.org/abs/1904.10520} {arXiv:1904.10520
  [hep-ex]} \BibitemShut {NoStop}%
\bibitem [{\citenamefont {Esser}\ \emph {et~al.}(2015)\citenamefont {Esser},
  \citenamefont {Nagao}, \citenamefont {Schulz}, \citenamefont {Achenbach},
  \citenamefont {Ayerbe~Gayoso}, \citenamefont {B\"ohm}, \citenamefont
  {Borodina}, \citenamefont {Bosnar}, \citenamefont {Bozkurt}, \citenamefont
  {Debenjak}, \citenamefont {Distler}, \citenamefont {Fri\ifmmode
  \check{s}\else \v{s}\fi{}\ifmmode \check{c}\else
  \v{c}\fi{}i\ifmmode~\acute{c}\else \'{c}\fi{}}, \citenamefont {Fujii},
  \citenamefont {Gogami}, \citenamefont {Hashimoto}, \citenamefont {Hirose},
  \citenamefont {Kanda}, \citenamefont {Kaneta}, \citenamefont {Kim},
  \citenamefont {Kohl}, \citenamefont {Kusaka}, \citenamefont {Margaryan},
  \citenamefont {Merkel}, \citenamefont {Mihovilovi\ifmmode~\check{c}\else
  \v{c}\fi{}}, \citenamefont {M\"uller}, \citenamefont {Nakamura},
  \citenamefont {Pochodzalla}, \citenamefont {Rappold}, \citenamefont
  {Reinhold}, \citenamefont {Saito}, \citenamefont {Sanchez~Lorente},
  \citenamefont {S\'anchez~Majos}, \citenamefont {Schlimme}, \citenamefont
  {Schoth}, \citenamefont {Sfienti}, \citenamefont {\ifmmode~\check{S}\else
  \v{S}\fi{}irca}, \citenamefont {Tang}, \citenamefont {Thiel}, \citenamefont
  {Tsukada}, \citenamefont {Weber},\ and\ \citenamefont {Yoshida}}]{Esser2014}%
  \BibitemOpen
  \bibfield  {author} {\bibinfo {author} {\bibfnamefont {A.}~\bibnamefont
  {Esser}}, \bibinfo {author} {\bibfnamefont {S.}~\bibnamefont {Nagao}},
  \bibinfo {author} {\bibfnamefont {F.}~\bibnamefont {Schulz}}, \bibinfo
  {author} {\bibfnamefont {P.}~\bibnamefont {Achenbach}}, \bibinfo {author}
  {\bibfnamefont {C.}~\bibnamefont {Ayerbe~Gayoso}}, \bibinfo {author}
  {\bibfnamefont {R.}~\bibnamefont {B\"ohm}}, \bibinfo {author} {\bibfnamefont
  {O.}~\bibnamefont {Borodina}}, \bibinfo {author} {\bibfnamefont
  {D.}~\bibnamefont {Bosnar}}, \bibinfo {author} {\bibfnamefont
  {V.}~\bibnamefont {Bozkurt}}, \bibinfo {author} {\bibfnamefont
  {L.}~\bibnamefont {Debenjak}}, \bibinfo {author} {\bibfnamefont {M.~O.}\
  \bibnamefont {Distler}}, \bibinfo {author} {\bibfnamefont {I.}~\bibnamefont
  {Fri\ifmmode \check{s}\else \v{s}\fi{}\ifmmode \check{c}\else
  \v{c}\fi{}i\ifmmode~\acute{c}\else \'{c}\fi{}}}, \bibinfo {author}
  {\bibfnamefont {Y.}~\bibnamefont {Fujii}}, \bibinfo {author} {\bibfnamefont
  {T.}~\bibnamefont {Gogami}}, \bibinfo {author} {\bibfnamefont
  {O.}~\bibnamefont {Hashimoto}}, \bibinfo {author} {\bibfnamefont
  {S.}~\bibnamefont {Hirose}}, \bibinfo {author} {\bibfnamefont
  {H.}~\bibnamefont {Kanda}}, \bibinfo {author} {\bibfnamefont
  {M.}~\bibnamefont {Kaneta}}, \bibinfo {author} {\bibfnamefont
  {E.}~\bibnamefont {Kim}}, \bibinfo {author} {\bibfnamefont {Y.}~\bibnamefont
  {Kohl}}, \bibinfo {author} {\bibfnamefont {J.}~\bibnamefont {Kusaka}},
  \bibinfo {author} {\bibfnamefont {A.}~\bibnamefont {Margaryan}}, \bibinfo
  {author} {\bibfnamefont {H.}~\bibnamefont {Merkel}}, \bibinfo {author}
  {\bibfnamefont {M.}~\bibnamefont {Mihovilovi\ifmmode~\check{c}\else
  \v{c}\fi{}}}, \bibinfo {author} {\bibfnamefont {U.}~\bibnamefont {M\"uller}},
  \bibinfo {author} {\bibfnamefont {S.~N.}\ \bibnamefont {Nakamura}}, \bibinfo
  {author} {\bibfnamefont {J.}~\bibnamefont {Pochodzalla}}, \bibinfo {author}
  {\bibfnamefont {C.}~\bibnamefont {Rappold}}, \bibinfo {author} {\bibfnamefont
  {J.}~\bibnamefont {Reinhold}}, \bibinfo {author} {\bibfnamefont {T.~R.}\
  \bibnamefont {Saito}}, \bibinfo {author} {\bibfnamefont {A.}~\bibnamefont
  {Sanchez~Lorente}}, \bibinfo {author} {\bibfnamefont {S.}~\bibnamefont
  {S\'anchez~Majos}}, \bibinfo {author} {\bibfnamefont {B.~S.}\ \bibnamefont
  {Schlimme}}, \bibinfo {author} {\bibfnamefont {M.}~\bibnamefont {Schoth}},
  \bibinfo {author} {\bibfnamefont {C.}~\bibnamefont {Sfienti}}, \bibinfo
  {author} {\bibfnamefont {S.}~\bibnamefont {\ifmmode~\check{S}\else
  \v{S}\fi{}irca}}, \bibinfo {author} {\bibfnamefont {L.}~\bibnamefont {Tang}},
  \bibinfo {author} {\bibfnamefont {M.}~\bibnamefont {Thiel}}, \bibinfo
  {author} {\bibfnamefont {K.}~\bibnamefont {Tsukada}}, \bibinfo {author}
  {\bibfnamefont {A.}~\bibnamefont {Weber}}, \ and\ \bibinfo {author}
  {\bibfnamefont {K.}~\bibnamefont {Yoshida}} (\bibinfo {collaboration} {A1
  Collaboration}),\ }\href {\doibase 10.1103/PhysRevLett.114.232501} {\bibfield
   {journal} {\bibinfo  {journal} {Phys. Rev. Lett.}\ }\textbf {\bibinfo
  {volume} {114}},\ \bibinfo {pages} {232501} (\bibinfo {year}
  {2015})}\BibitemShut {NoStop}%
\bibitem [{\citenamefont {Yamamoto}\ \emph {et~al.}(2015)\citenamefont
  {Yamamoto}, \citenamefont {Agnello}, \citenamefont {Akazawa}, \citenamefont
  {Amano}, \citenamefont {Aoki}, \citenamefont {Botta}, \citenamefont {Chiga},
  \citenamefont {Ekawa}, \citenamefont {Evtoukhovitch}, \citenamefont
  {Feliciello}, \citenamefont {Fujita}, \citenamefont {Gogami}, \citenamefont
  {Hasegawa}, \citenamefont {Hayakawa}, \citenamefont {Hayakawa}, \citenamefont
  {Honda}, \citenamefont {Hosomi}, \citenamefont {Hwang}, \citenamefont
  {Ichige}, \citenamefont {Ichikawa}, \citenamefont {Ikeda}, \citenamefont
  {Imai}, \citenamefont {Ishimoto}, \citenamefont {Kanatsuki}, \citenamefont
  {Kim}, \citenamefont {Kim}, \citenamefont {Kinbara}, \citenamefont {Koike},
  \citenamefont {Lee}, \citenamefont {Marcello}, \citenamefont {Miwa},
  \citenamefont {Moon}, \citenamefont {Nagae}, \citenamefont {Nagao},
  \citenamefont {Nakada}, \citenamefont {Nakagawa}, \citenamefont {Ogura},
  \citenamefont {Sakaguchi}, \citenamefont {Sako}, \citenamefont {Sasaki},
  \citenamefont {Sato}, \citenamefont {Shiozaki}, \citenamefont {Shirotori},
  \citenamefont {Sugimura}, \citenamefont {Suto}, \citenamefont {Suzuki},
  \citenamefont {Takahashi}, \citenamefont {Tamura}, \citenamefont {Tanabe},
  \citenamefont {Tanida}, \citenamefont {Tsamalaidze}, \citenamefont {Ukai},
  \citenamefont {Yamamoto},\ and\ \citenamefont {Yang}}]{Yamamoto2015}%
  \BibitemOpen
  \bibfield  {author} {\bibinfo {author} {\bibfnamefont {T.~O.}\ \bibnamefont
  {Yamamoto}}, \bibinfo {author} {\bibfnamefont {M.}~\bibnamefont {Agnello}},
  \bibinfo {author} {\bibfnamefont {Y.}~\bibnamefont {Akazawa}}, \bibinfo
  {author} {\bibfnamefont {N.}~\bibnamefont {Amano}}, \bibinfo {author}
  {\bibfnamefont {K.}~\bibnamefont {Aoki}}, \bibinfo {author} {\bibfnamefont
  {E.}~\bibnamefont {Botta}}, \bibinfo {author} {\bibfnamefont
  {N.}~\bibnamefont {Chiga}}, \bibinfo {author} {\bibfnamefont
  {H.}~\bibnamefont {Ekawa}}, \bibinfo {author} {\bibfnamefont
  {P.}~\bibnamefont {Evtoukhovitch}}, \bibinfo {author} {\bibfnamefont
  {A.}~\bibnamefont {Feliciello}}, \bibinfo {author} {\bibfnamefont
  {M.}~\bibnamefont {Fujita}}, \bibinfo {author} {\bibfnamefont
  {T.}~\bibnamefont {Gogami}}, \bibinfo {author} {\bibfnamefont
  {S.}~\bibnamefont {Hasegawa}}, \bibinfo {author} {\bibfnamefont {S.~H.}\
  \bibnamefont {Hayakawa}}, \bibinfo {author} {\bibfnamefont {T.}~\bibnamefont
  {Hayakawa}}, \bibinfo {author} {\bibfnamefont {R.}~\bibnamefont {Honda}},
  \bibinfo {author} {\bibfnamefont {K.}~\bibnamefont {Hosomi}}, \bibinfo
  {author} {\bibfnamefont {S.~H.}\ \bibnamefont {Hwang}}, \bibinfo {author}
  {\bibfnamefont {N.}~\bibnamefont {Ichige}}, \bibinfo {author} {\bibfnamefont
  {Y.}~\bibnamefont {Ichikawa}}, \bibinfo {author} {\bibfnamefont
  {M.}~\bibnamefont {Ikeda}}, \bibinfo {author} {\bibfnamefont
  {K.}~\bibnamefont {Imai}}, \bibinfo {author} {\bibfnamefont {S.}~\bibnamefont
  {Ishimoto}}, \bibinfo {author} {\bibfnamefont {S.}~\bibnamefont {Kanatsuki}},
  \bibinfo {author} {\bibfnamefont {M.~H.}\ \bibnamefont {Kim}}, \bibinfo
  {author} {\bibfnamefont {S.~H.}\ \bibnamefont {Kim}}, \bibinfo {author}
  {\bibfnamefont {S.}~\bibnamefont {Kinbara}}, \bibinfo {author} {\bibfnamefont
  {T.}~\bibnamefont {Koike}}, \bibinfo {author} {\bibfnamefont {J.~Y.}\
  \bibnamefont {Lee}}, \bibinfo {author} {\bibfnamefont {S.}~\bibnamefont
  {Marcello}}, \bibinfo {author} {\bibfnamefont {K.}~\bibnamefont {Miwa}},
  \bibinfo {author} {\bibfnamefont {T.}~\bibnamefont {Moon}}, \bibinfo {author}
  {\bibfnamefont {T.}~\bibnamefont {Nagae}}, \bibinfo {author} {\bibfnamefont
  {S.}~\bibnamefont {Nagao}}, \bibinfo {author} {\bibfnamefont
  {Y.}~\bibnamefont {Nakada}}, \bibinfo {author} {\bibfnamefont
  {M.}~\bibnamefont {Nakagawa}}, \bibinfo {author} {\bibfnamefont
  {Y.}~\bibnamefont {Ogura}}, \bibinfo {author} {\bibfnamefont
  {A.}~\bibnamefont {Sakaguchi}}, \bibinfo {author} {\bibfnamefont
  {H.}~\bibnamefont {Sako}}, \bibinfo {author} {\bibfnamefont {Y.}~\bibnamefont
  {Sasaki}}, \bibinfo {author} {\bibfnamefont {S.}~\bibnamefont {Sato}},
  \bibinfo {author} {\bibfnamefont {T.}~\bibnamefont {Shiozaki}}, \bibinfo
  {author} {\bibfnamefont {K.}~\bibnamefont {Shirotori}}, \bibinfo {author}
  {\bibfnamefont {H.}~\bibnamefont {Sugimura}}, \bibinfo {author}
  {\bibfnamefont {S.}~\bibnamefont {Suto}}, \bibinfo {author} {\bibfnamefont
  {S.}~\bibnamefont {Suzuki}}, \bibinfo {author} {\bibfnamefont
  {T.}~\bibnamefont {Takahashi}}, \bibinfo {author} {\bibfnamefont
  {H.}~\bibnamefont {Tamura}}, \bibinfo {author} {\bibfnamefont
  {K.}~\bibnamefont {Tanabe}}, \bibinfo {author} {\bibfnamefont
  {K.}~\bibnamefont {Tanida}}, \bibinfo {author} {\bibfnamefont
  {Z.}~\bibnamefont {Tsamalaidze}}, \bibinfo {author} {\bibfnamefont
  {M.}~\bibnamefont {Ukai}}, \bibinfo {author} {\bibfnamefont {Y.}~\bibnamefont
  {Yamamoto}}, \ and\ \bibinfo {author} {\bibfnamefont {S.~B.}\ \bibnamefont
  {Yang}} (\bibinfo {collaboration} {J-PARC E13 Collaboration}),\ }\href
  {\doibase 10.1103/PhysRevLett.115.222501} {\bibfield  {journal} {\bibinfo
  {journal} {Phys. Rev. Lett.}\ }\textbf {\bibinfo {volume} {115}},\ \bibinfo
  {pages} {222501} (\bibinfo {year} {2015})}\BibitemShut {NoStop}%
\bibitem [{\citenamefont {Braun-Munzinger}\ and\ \citenamefont
  {D\"onigus}(2019)}]{Braun-Munzinger2018}%
  \BibitemOpen
  \bibfield  {author} {\bibinfo {author} {\bibfnamefont {P.}~\bibnamefont
  {Braun-Munzinger}}\ and\ \bibinfo {author} {\bibfnamefont {B.}~\bibnamefont
  {D\"onigus}},\ }\href {\doibase 10.1016/j.nuclphysa.2019.02.006} {\bibfield
  {journal} {\bibinfo  {journal} {Nucl. Phys. A}\ }\textbf {\bibinfo {volume}
  {987}},\ \bibinfo {pages} {144} (\bibinfo {year} {2019})},\ \Eprint
  {http://arxiv.org/abs/1809.04681} {arXiv:1809.04681 [nucl-ex]} \BibitemShut
  {NoStop}%
\bibitem [{\citenamefont {Qin}\ \emph {et~al.}(2012)\citenamefont {Qin} \emph
  {et~al.}}]{Qin2011}%
  \BibitemOpen
  \bibfield  {author} {\bibinfo {author} {\bibfnamefont {L.}~\bibnamefont
  {Qin}} \emph {et~al.},\ }\href {\doibase 10.1103/PhysRevLett.108.172701}
  {\bibfield  {journal} {\bibinfo  {journal} {Phys. Rev. Lett.}\ }\textbf
  {\bibinfo {volume} {108}},\ \bibinfo {pages} {172701} (\bibinfo {year}
  {2012})},\ \Eprint {http://arxiv.org/abs/1110.3345} {arXiv:1110.3345
  [nucl-ex]} \BibitemShut {NoStop}%
\bibitem [{\citenamefont {Bougault}\ \emph {et~al.}(2020)\citenamefont
  {Bougault} \emph {et~al.}}]{indra}%
  \BibitemOpen
  \bibfield  {author} {\bibinfo {author} {\bibfnamefont {R.}~\bibnamefont
  {Bougault}} \emph {et~al.},\ }\href {\doibase 10.1088/1361-6471/ab56ba}
  {\bibfield  {journal} {\bibinfo  {journal} {J. Phys. G}\ }\textbf {\bibinfo
  {volume} {47}},\ \bibinfo {pages} {025103} (\bibinfo {year} {2020})},\
  \Eprint {http://arxiv.org/abs/1911.08355} {arXiv:1911.08355 [nucl-ex]}
  \BibitemShut {NoStop}%
\bibitem [{\citenamefont {Arcones}\ \emph {et~al.}(2008)\citenamefont
  {Arcones}, \citenamefont {Martinez-Pinedo}, \citenamefont {O'Connor},
  \citenamefont {Schwenk}, \citenamefont {Janka}, \citenamefont {Horowitz},\
  and\ \citenamefont {Langanke}}]{Arcones2008}%
  \BibitemOpen
  \bibfield  {author} {\bibinfo {author} {\bibfnamefont {A.}~\bibnamefont
  {Arcones}}, \bibinfo {author} {\bibfnamefont {G.}~\bibnamefont
  {Martinez-Pinedo}}, \bibinfo {author} {\bibfnamefont {E.}~\bibnamefont
  {O'Connor}}, \bibinfo {author} {\bibfnamefont {A.}~\bibnamefont {Schwenk}},
  \bibinfo {author} {\bibfnamefont {H.~T.}\ \bibnamefont {Janka}}, \bibinfo
  {author} {\bibfnamefont {C.~J.}\ \bibnamefont {Horowitz}}, \ and\ \bibinfo
  {author} {\bibfnamefont {K.}~\bibnamefont {Langanke}},\ }\href {\doibase
  10.1103/PhysRevC.78.015806} {\bibfield  {journal} {\bibinfo  {journal} {Phys.
  Rev. C}\ }\textbf {\bibinfo {volume} {78}},\ \bibinfo {pages} {015806}
  (\bibinfo {year} {2008})},\ \Eprint {http://arxiv.org/abs/0805.3752}
  {arXiv:0805.3752 [astro-ph]} \BibitemShut {NoStop}%
\bibitem [{\citenamefont {Fischer}\ \emph {et~al.}(2020)\citenamefont
  {Fischer}, \citenamefont {Typel}, \citenamefont {R\"opke}, \citenamefont
  {Bastian},\ and\ \citenamefont {Mart\'\i{}nez-Pinedo}}]{Fischer2020}%
  \BibitemOpen
  \bibfield  {author} {\bibinfo {author} {\bibfnamefont {T.}~\bibnamefont
  {Fischer}}, \bibinfo {author} {\bibfnamefont {S.}~\bibnamefont {Typel}},
  \bibinfo {author} {\bibfnamefont {G.}~\bibnamefont {R\"opke}}, \bibinfo
  {author} {\bibfnamefont {N.-U.~F.}\ \bibnamefont {Bastian}}, \ and\ \bibinfo
  {author} {\bibfnamefont {G.}~\bibnamefont {Mart\'\i{}nez-Pinedo}},\ }\href
  {\doibase 10.1103/PhysRevC.102.055807} {\bibfield  {journal} {\bibinfo
  {journal} {Phys. Rev. C}\ }\textbf {\bibinfo {volume} {102}},\ \bibinfo
  {pages} {055807} (\bibinfo {year} {2020})},\ \Eprint
  {http://arxiv.org/abs/2008.13608} {arXiv:2008.13608 [astro-ph.HE]}
  \BibitemShut {NoStop}%
\bibitem [{\citenamefont {Rosswog}(2015)}]{Rosswog2015}%
  \BibitemOpen
  \bibfield  {author} {\bibinfo {author} {\bibfnamefont {S.}~\bibnamefont
  {Rosswog}},\ }\href {\doibase 10.1142/S0218271815300128} {\bibfield
  {journal} {\bibinfo  {journal} {Int. J. Mod. Phys. D}\ }\textbf {\bibinfo
  {volume} {24}},\ \bibinfo {pages} {1530012} (\bibinfo {year} {2015})},\
  \Eprint {http://arxiv.org/abs/1501.02081} {arXiv:1501.02081 [astro-ph.HE]}
  \BibitemShut {NoStop}%
\bibitem [{\citenamefont {Alford}\ \emph {et~al.}(2018)\citenamefont {Alford},
  \citenamefont {Bovard}, \citenamefont {Hanauske}, \citenamefont {Rezzolla},\
  and\ \citenamefont {Schwenzer}}]{Alford2018}%
  \BibitemOpen
  \bibfield  {author} {\bibinfo {author} {\bibfnamefont {M.~G.}\ \bibnamefont
  {Alford}}, \bibinfo {author} {\bibfnamefont {L.}~\bibnamefont {Bovard}},
  \bibinfo {author} {\bibfnamefont {M.}~\bibnamefont {Hanauske}}, \bibinfo
  {author} {\bibfnamefont {L.}~\bibnamefont {Rezzolla}}, \ and\ \bibinfo
  {author} {\bibfnamefont {K.}~\bibnamefont {Schwenzer}},\ }\href {\doibase
  10.1103/PhysRevLett.120.041101} {\bibfield  {journal} {\bibinfo  {journal}
  {Phys. Rev. Lett.}\ }\textbf {\bibinfo {volume} {120}},\ \bibinfo {pages}
  {041101} (\bibinfo {year} {2018})},\ \Eprint
  {http://arxiv.org/abs/1707.09475} {arXiv:1707.09475 [gr-qc]} \BibitemShut
  {NoStop}%
\bibitem [{\citenamefont {Fujibayashi}\ \emph {et~al.}(2018)\citenamefont
  {Fujibayashi}, \citenamefont {Kiuchi}, \citenamefont {Nishimura},
  \citenamefont {Sekiguchi},\ and\ \citenamefont {Shibata}}]{Fujibayashi2017}%
  \BibitemOpen
  \bibfield  {author} {\bibinfo {author} {\bibfnamefont {S.}~\bibnamefont
  {Fujibayashi}}, \bibinfo {author} {\bibfnamefont {K.}~\bibnamefont {Kiuchi}},
  \bibinfo {author} {\bibfnamefont {N.}~\bibnamefont {Nishimura}}, \bibinfo
  {author} {\bibfnamefont {Y.}~\bibnamefont {Sekiguchi}}, \ and\ \bibinfo
  {author} {\bibfnamefont {M.}~\bibnamefont {Shibata}},\ }\href {\doibase
  10.3847/1538-4357/aabafd} {\bibfield  {journal} {\bibinfo  {journal}
  {Astrophys. J.}\ }\textbf {\bibinfo {volume} {860}},\ \bibinfo {pages} {64}
  (\bibinfo {year} {2018})},\ \Eprint {http://arxiv.org/abs/1711.02093}
  {arXiv:1711.02093 [astro-ph.HE]} \BibitemShut {NoStop}%
\bibitem [{\citenamefont {Glendenning}(2000)}]{Glend}%
  \BibitemOpen
  \bibfield  {author} {\bibinfo {author} {\bibfnamefont {N.~K.}\ \bibnamefont
  {Glendenning}},\ }\href@noop {} {\emph {\bibinfo {title} {Compact Stars:
  Nuclear Physics, Particle Physics, and General Relativity (Astron. Astrophys.
  Library)}}}\ (\bibinfo  {publisher} {Springer},\ \bibinfo {year}
  {2000})\BibitemShut {NoStop}%
\bibitem [{\citenamefont {Marques}\ \emph {et~al.}(2017)\citenamefont
  {Marques}, \citenamefont {Oertel}, \citenamefont {Hempel},\ and\
  \citenamefont {Novak}}]{Marques2017}%
  \BibitemOpen
  \bibfield  {author} {\bibinfo {author} {\bibfnamefont {M.}~\bibnamefont
  {Marques}}, \bibinfo {author} {\bibfnamefont {M.}~\bibnamefont {Oertel}},
  \bibinfo {author} {\bibfnamefont {M.}~\bibnamefont {Hempel}}, \ and\ \bibinfo
  {author} {\bibfnamefont {J.}~\bibnamefont {Novak}},\ }\href {\doibase
  10.1103/PhysRevC.96.045806} {\bibfield  {journal} {\bibinfo  {journal} {Phys.
  Rev. C}\ }\textbf {\bibinfo {volume} {96}},\ \bibinfo {pages} {045806}
  (\bibinfo {year} {2017})},\ \Eprint {http://arxiv.org/abs/1706.02913}
  {arXiv:1706.02913 [nucl-th]} \BibitemShut {NoStop}%
\bibitem [{\citenamefont {Fortin}\ \emph {et~al.}(2018)\citenamefont {Fortin},
  \citenamefont {Oertel},\ and\ \citenamefont {Provid\^encia}}]{Fortin2017}%
  \BibitemOpen
  \bibfield  {author} {\bibinfo {author} {\bibfnamefont {M.}~\bibnamefont
  {Fortin}}, \bibinfo {author} {\bibfnamefont {M.}~\bibnamefont {Oertel}}, \
  and\ \bibinfo {author} {\bibfnamefont {C.}~\bibnamefont {Provid\^encia}},\
  }\href {\doibase 10.1017/pasa.2018.32} {\bibfield  {journal} {\bibinfo
  {journal} {Publ. Astron. Soc. Austral.}\ }\textbf {\bibinfo {volume} {35}},\
  \bibinfo {pages} {44} (\bibinfo {year} {2018})},\ \Eprint
  {http://arxiv.org/abs/1711.09427} {arXiv:1711.09427 [astro-ph.HE]}
  \BibitemShut {NoStop}%
\bibitem [{\citenamefont {Typel}\ \emph {et~al.}(2010)\citenamefont {Typel},
  \citenamefont {Ropke}, \citenamefont {Klahn}, \citenamefont {Blaschke},\ and\
  \citenamefont {Wolter}}]{Typel2009}%
  \BibitemOpen
  \bibfield  {author} {\bibinfo {author} {\bibfnamefont {S.}~\bibnamefont
  {Typel}}, \bibinfo {author} {\bibfnamefont {G.}~\bibnamefont {Ropke}},
  \bibinfo {author} {\bibfnamefont {T.}~\bibnamefont {Klahn}}, \bibinfo
  {author} {\bibfnamefont {D.}~\bibnamefont {Blaschke}}, \ and\ \bibinfo
  {author} {\bibfnamefont {H.~H.}\ \bibnamefont {Wolter}},\ }\href {\doibase
  10.1103/PhysRevC.81.015803} {\bibfield  {journal} {\bibinfo  {journal} {Phys.
  Rev. C}\ }\textbf {\bibinfo {volume} {81}},\ \bibinfo {pages} {015803}
  (\bibinfo {year} {2010})},\ \Eprint {http://arxiv.org/abs/0908.2344}
  {arXiv:0908.2344 [nucl-th]} \BibitemShut {NoStop}%
\bibitem [{\citenamefont {Steiner}\ \emph {et~al.}(2013)\citenamefont
  {Steiner}, \citenamefont {Hempel},\ and\ \citenamefont
  {Fischer}}]{Steiner2012}%
  \BibitemOpen
  \bibfield  {author} {\bibinfo {author} {\bibfnamefont {A.~W.}\ \bibnamefont
  {Steiner}}, \bibinfo {author} {\bibfnamefont {M.}~\bibnamefont {Hempel}}, \
  and\ \bibinfo {author} {\bibfnamefont {T.}~\bibnamefont {Fischer}},\ }\href
  {\doibase 10.1088/0004-637X/774/1/17} {\bibfield  {journal} {\bibinfo
  {journal} {Astrophys. J.}\ }\textbf {\bibinfo {volume} {774}},\ \bibinfo
  {pages} {17} (\bibinfo {year} {2013})},\ \Eprint
  {http://arxiv.org/abs/1207.2184} {arXiv:1207.2184 [astro-ph.SR]} \BibitemShut
  {NoStop}%
\bibitem [{Note1()}]{Note1}%
  \BibitemOpen
  \bibinfo {note} {Url{https://compose.obspm.fr/}}\BibitemShut {NoStop}%
\bibitem [{\citenamefont {Lattimer}\ and\ \citenamefont {Swesty}(1991)}]{LS}%
  \BibitemOpen
  \bibfield  {author} {\bibinfo {author} {\bibfnamefont {J.~M.}\ \bibnamefont
  {Lattimer}}\ and\ \bibinfo {author} {\bibfnamefont {F.~D.}\ \bibnamefont
  {Swesty}},\ }\href {\doibase 10.1016/0375-9474(91)90452-C} {\bibfield
  {journal} {\bibinfo  {journal} {Nucl. Phys. A}\ }\textbf {\bibinfo {volume}
  {535}},\ \bibinfo {pages} {331} (\bibinfo {year} {1991})}\BibitemShut
  {NoStop}%
\bibitem [{\citenamefont {Shen}\ \emph {et~al.}(1998)\citenamefont {Shen},
  \citenamefont {Toki}, \citenamefont {Oyamatsu},\ and\ \citenamefont
  {Sumiyoshi}}]{Shen1998}%
  \BibitemOpen
  \bibfield  {author} {\bibinfo {author} {\bibfnamefont {H.}~\bibnamefont
  {Shen}}, \bibinfo {author} {\bibfnamefont {H.}~\bibnamefont {Toki}}, \bibinfo
  {author} {\bibfnamefont {K.}~\bibnamefont {Oyamatsu}}, \ and\ \bibinfo
  {author} {\bibfnamefont {K.}~\bibnamefont {Sumiyoshi}},\ }\href {\doibase
  10.1016/S0375-9474(98)00236-X} {\bibfield  {journal} {\bibinfo  {journal}
  {Nucl. Phys. A}\ }\textbf {\bibinfo {volume} {637}},\ \bibinfo {pages} {435}
  (\bibinfo {year} {1998})},\ \Eprint {http://arxiv.org/abs/nucl-th/9805035}
  {arXiv:nucl-th/9805035} \BibitemShut {NoStop}%
\bibitem [{\citenamefont {Hempel}\ and\ \citenamefont
  {Schaffner-Bielich}(2010)}]{Hempel2010}%
  \BibitemOpen
  \bibfield  {author} {\bibinfo {author} {\bibfnamefont {M.}~\bibnamefont
  {Hempel}}\ and\ \bibinfo {author} {\bibfnamefont {J.}~\bibnamefont
  {Schaffner-Bielich}},\ }\href {\doibase 10.1016/j.nuclphysa.2010.02.010}
  {\bibfield  {journal} {\bibinfo  {journal} {Nucl. Phys. A}\ }\textbf
  {\bibinfo {volume} {837}},\ \bibinfo {pages} {210} (\bibinfo {year}
  {2010})},\ \Eprint {http://arxiv.org/abs/0911.4073} {arXiv:0911.4073
  [nucl-th]} \BibitemShut {NoStop}%
\bibitem [{\citenamefont {Raduta}\ and\ \citenamefont
  {Gulminelli}(2010)}]{Raduta2010}%
  \BibitemOpen
  \bibfield  {author} {\bibinfo {author} {\bibfnamefont {A.~R.}\ \bibnamefont
  {Raduta}}\ and\ \bibinfo {author} {\bibfnamefont {F.}~\bibnamefont
  {Gulminelli}},\ }\href {\doibase 10.1103/PhysRevC.82.065801} {\bibfield
  {journal} {\bibinfo  {journal} {Phys. Rev. C}\ }\textbf {\bibinfo {volume}
  {82}},\ \bibinfo {pages} {065801} (\bibinfo {year} {2010})},\ \Eprint
  {http://arxiv.org/abs/1009.2226} {arXiv:1009.2226 [nucl-th]} \BibitemShut
  {NoStop}%
\bibitem [{\citenamefont {Avancini}\ \emph {et~al.}(2010)\citenamefont
  {Avancini}, \citenamefont {Barros}, \citenamefont {Menezes},\ and\
  \citenamefont {Providencia}}]{Avancini2010}%
  \BibitemOpen
  \bibfield  {author} {\bibinfo {author} {\bibfnamefont {S.~S.}\ \bibnamefont
  {Avancini}}, \bibinfo {author} {\bibfnamefont {C.~C.}\ \bibnamefont {Barros},
  \bibfnamefont {Jr.}}, \bibinfo {author} {\bibfnamefont {D.~P.}\ \bibnamefont
  {Menezes}}, \ and\ \bibinfo {author} {\bibfnamefont {C.}~\bibnamefont
  {Providencia}},\ }\href {\doibase 10.1103/PhysRevC.82.025808} {\bibfield
  {journal} {\bibinfo  {journal} {Phys. Rev. C}\ }\textbf {\bibinfo {volume}
  {82}},\ \bibinfo {pages} {025808} (\bibinfo {year} {2010})},\ \Eprint
  {http://arxiv.org/abs/1007.2319} {arXiv:1007.2319 [nucl-th]} \BibitemShut
  {NoStop}%
\bibitem [{\citenamefont {Ferreira}\ and\ \citenamefont
  {Providencia}(2012)}]{Ferreira2012}%
  \BibitemOpen
  \bibfield  {author} {\bibinfo {author} {\bibfnamefont {M.}~\bibnamefont
  {Ferreira}}\ and\ \bibinfo {author} {\bibfnamefont {C.}~\bibnamefont
  {Providencia}},\ }\href {\doibase 10.1103/PhysRevC.85.055811} {\bibfield
  {journal} {\bibinfo  {journal} {Phys. Rev. C}\ }\textbf {\bibinfo {volume}
  {85}},\ \bibinfo {pages} {055811} (\bibinfo {year} {2012})},\ \Eprint
  {http://arxiv.org/abs/1206.0139} {arXiv:1206.0139 [nucl-th]} \BibitemShut
  {NoStop}%
\bibitem [{\citenamefont {Ropke}(2009)}]{Ropke2008}%
  \BibitemOpen
  \bibfield  {author} {\bibinfo {author} {\bibfnamefont {G.}~\bibnamefont
  {Ropke}},\ }\href {\doibase 10.1103/PhysRevC.79.014002} {\bibfield  {journal}
  {\bibinfo  {journal} {Phys. Rev. C}\ }\textbf {\bibinfo {volume} {79}},\
  \bibinfo {pages} {014002} (\bibinfo {year} {2009})},\ \Eprint
  {http://arxiv.org/abs/0810.4645} {arXiv:0810.4645 [nucl-th]} \BibitemShut
  {NoStop}%
\bibitem [{\citenamefont {Ropke}(2011)}]{Ropke2011}%
  \BibitemOpen
  \bibfield  {author} {\bibinfo {author} {\bibfnamefont {G.}~\bibnamefont
  {Ropke}},\ }\href {\doibase 10.1016/j.nuclphysa.2011.07.010} {\bibfield
  {journal} {\bibinfo  {journal} {Nucl. Phys. A}\ }\textbf {\bibinfo {volume}
  {867}},\ \bibinfo {pages} {66} (\bibinfo {year} {2011})},\ \Eprint
  {http://arxiv.org/abs/1101.4685} {arXiv:1101.4685 [nucl-th]} \BibitemShut
  {NoStop}%
\bibitem [{\citenamefont {R\"opke}(2015)}]{Ropke2015}%
  \BibitemOpen
  \bibfield  {author} {\bibinfo {author} {\bibfnamefont {G.}~\bibnamefont
  {R\"opke}},\ }\href {\doibase 10.1103/PhysRevC.92.054001} {\bibfield
  {journal} {\bibinfo  {journal} {Phys. Rev. C}\ }\textbf {\bibinfo {volume}
  {92}},\ \bibinfo {pages} {054001} (\bibinfo {year} {2015})},\ \Eprint
  {http://arxiv.org/abs/1411.4593} {arXiv:1411.4593 [nucl-th]} \BibitemShut
  {NoStop}%
\bibitem [{\citenamefont {R\"opke}(2020)}]{Ropke2020}%
  \BibitemOpen
  \bibfield  {author} {\bibinfo {author} {\bibfnamefont {G.}~\bibnamefont
  {R\"opke}},\ }\href {\doibase 10.1103/PhysRevC.101.064310} {\bibfield
  {journal} {\bibinfo  {journal} {Phys. Rev. C}\ }\textbf {\bibinfo {volume}
  {101}},\ \bibinfo {pages} {064310} (\bibinfo {year} {2020})},\ \Eprint
  {http://arxiv.org/abs/2004.09773} {arXiv:2004.09773 [nucl-th]} \BibitemShut
  {NoStop}%
\bibitem [{\citenamefont {Pais}\ \emph {et~al.}(2018)\citenamefont {Pais},
  \citenamefont {Gulminelli}, \citenamefont {Provid\^encia},\ and\
  \citenamefont {R\"opke}}]{Pais2018}%
  \BibitemOpen
  \bibfield  {author} {\bibinfo {author} {\bibfnamefont {H.}~\bibnamefont
  {Pais}}, \bibinfo {author} {\bibfnamefont {F.}~\bibnamefont {Gulminelli}},
  \bibinfo {author} {\bibfnamefont {C.}~\bibnamefont {Provid\^encia}}, \ and\
  \bibinfo {author} {\bibfnamefont {G.}~\bibnamefont {R\"opke}},\ }\href
  {\doibase 10.1103/PhysRevC.97.045805} {\bibfield  {journal} {\bibinfo
  {journal} {Phys. Rev. C}\ }\textbf {\bibinfo {volume} {97}},\ \bibinfo
  {pages} {045805} (\bibinfo {year} {2018})},\ \Eprint
  {http://arxiv.org/abs/1804.01328} {arXiv:1804.01328 [nucl-th]} \BibitemShut
  {NoStop}%
\bibitem [{\citenamefont {Pais}\ \emph
  {et~al.}(2020{\natexlab{a}})\citenamefont {Pais} \emph {et~al.}}]{Pais2020prl}%
  \BibitemOpen
  \bibfield  {author} {\bibinfo {author} {\bibfnamefont {H.}~\bibnamefont
  {Pais}} \emph {et~al.},\ }\href {\doibase 10.1103/PhysRevLett.125.012701}
  {\bibfield  {journal} {\bibinfo  {journal} {Phys. Rev. Lett.}\ }\textbf
  {\bibinfo {volume} {125}},\ \bibinfo {pages} {012701} (\bibinfo {year}
  {2020}{\natexlab{a}})},\ \Eprint {http://arxiv.org/abs/1911.10849}
  {arXiv:1911.10849 [nucl-th]} \BibitemShut {NoStop}%
\bibitem [{\citenamefont {Pais}\ \emph
  {et~al.}(2020{\natexlab{b}})\citenamefont {Pais} \emph {et~al.}}]{Pais2020}%
  \BibitemOpen
  \bibfield  {author} {\bibinfo {author} {\bibfnamefont {H.}~\bibnamefont
  {Pais}} \emph {et~al.},\ }\href {\doibase 10.1088/1361-6471/aba561}
  {\bibfield  {journal} {\bibinfo  {journal} {J. Phys. G}\ }\textbf {\bibinfo
  {volume} {47}},\ \bibinfo {pages} {105204} (\bibinfo {year}
  {2020}{\natexlab{b}})},\ \Eprint {http://arxiv.org/abs/2006.07256}
  {arXiv:2006.07256 [nucl-th]} \BibitemShut {NoStop}%
\bibitem [{\citenamefont {Cust\'odio}\ \emph {et~al.}(2020)\citenamefont
  {Cust\'odio}, \citenamefont {Falc\~ao}, \citenamefont {Pais}, \citenamefont
  {Provid\^encia}, \citenamefont {Gulminelli},\ and\ \citenamefont
  {R\"opke}}]{Custodio2020}%
  \BibitemOpen
  \bibfield  {author} {\bibinfo {author} {\bibfnamefont {T.}~\bibnamefont
  {Cust\'odio}}, \bibinfo {author} {\bibfnamefont {A.}~\bibnamefont
  {Falc\~ao}}, \bibinfo {author} {\bibfnamefont {H.}~\bibnamefont {Pais}},
  \bibinfo {author} {\bibfnamefont {C.}~\bibnamefont {Provid\^encia}}, \bibinfo
  {author} {\bibfnamefont {F.}~\bibnamefont {Gulminelli}}, \ and\ \bibinfo
  {author} {\bibfnamefont {G.}~\bibnamefont {R\"opke}},\ }\href {\doibase
  10.1140/epja/s10050-020-00302-w} {\bibfield  {journal} {\bibinfo  {journal}
  {Eur. Phys. J. A}\ }\textbf {\bibinfo {volume} {56}},\ \bibinfo {pages} {295}
  (\bibinfo {year} {2020})},\ \Eprint {http://arxiv.org/abs/2009.14035}
  {arXiv:2009.14035 [nucl-th]} \BibitemShut {NoStop}%
\bibitem [{\citenamefont {Menezes}\ and\ \citenamefont
  {Provid\^encia}(2017)}]{Menezes2017}%
  \BibitemOpen
  \bibfield  {author} {\bibinfo {author} {\bibfnamefont {D.~P.}\ \bibnamefont
  {Menezes}}\ and\ \bibinfo {author} {\bibfnamefont {C.}~\bibnamefont
  {Provid\^encia}},\ }\href {\doibase 10.1103/PhysRevC.96.045803} {\bibfield
  {journal} {\bibinfo  {journal} {Phys. Rev. C}\ }\textbf {\bibinfo {volume}
  {96}},\ \bibinfo {pages} {045803} (\bibinfo {year} {2017})},\ \Eprint
  {http://arxiv.org/abs/1707.01338} {arXiv:1707.01338 [nucl-th]} \BibitemShut
  {NoStop}%
\bibitem [{\citenamefont {Sedrakian}(2020)}]{Sedrakian2020}%
  \BibitemOpen
  \bibfield  {author} {\bibinfo {author} {\bibfnamefont {A.}~\bibnamefont
  {Sedrakian}},\ }\href {\doibase 10.1140/epja/s10050-020-00262-1} {\bibfield
  {journal} {\bibinfo  {journal} {Eur. Phys. J. A}\ }\textbf {\bibinfo {volume}
  {56}},\ \bibinfo {pages} {258} (\bibinfo {year} {2020})},\ \Eprint
  {http://arxiv.org/abs/2009.00357} {arXiv:2009.00357 [nucl-th]} \BibitemShut
  {NoStop}%
\bibitem [{\citenamefont {Weissenborn}\ \emph {et~al.}(2012)\citenamefont
  {Weissenborn}, \citenamefont {Chatterjee},\ and\ \citenamefont
  {Schaffner-Bielich}}]{Weissenborn2011}%
  \BibitemOpen
  \bibfield  {author} {\bibinfo {author} {\bibfnamefont {S.}~\bibnamefont
  {Weissenborn}}, \bibinfo {author} {\bibfnamefont {D.}~\bibnamefont
  {Chatterjee}}, \ and\ \bibinfo {author} {\bibfnamefont {J.}~\bibnamefont
  {Schaffner-Bielich}},\ }\href {\doibase 10.1103/PhysRevC.85.065802}
  {\bibfield  {journal} {\bibinfo  {journal} {Phys. Rev. C}\ }\textbf {\bibinfo
  {volume} {85}},\ \bibinfo {pages} {065802} (\bibinfo {year} {2012})},\
  \bibinfo {note} {[Erratum: Phys.Rev.C 90, 019904 (2014)]},\ \Eprint
  {http://arxiv.org/abs/1112.0234} {arXiv:1112.0234 [astro-ph.HE]} \BibitemShut
  {NoStop}%
\bibitem [{\citenamefont {Fortin}\ \emph {et~al.}(2020)\citenamefont {Fortin},
  \citenamefont {Raduta}, \citenamefont {Avancini},\ and\ \citenamefont
  {Provid\^encia}}]{Fortin2020}%
  \BibitemOpen
  \bibfield  {author} {\bibinfo {author} {\bibfnamefont {M.}~\bibnamefont
  {Fortin}}, \bibinfo {author} {\bibfnamefont {A.~R.}\ \bibnamefont {Raduta}},
  \bibinfo {author} {\bibfnamefont {S.}~\bibnamefont {Avancini}}, \ and\
  \bibinfo {author} {\bibfnamefont {C.}~\bibnamefont {Provid\^encia}},\ }\href
  {\doibase 10.1103/PhysRevD.101.034017} {\bibfield  {journal} {\bibinfo
  {journal} {Phys. Rev. D}\ }\textbf {\bibinfo {volume} {101}},\ \bibinfo
  {pages} {034017} (\bibinfo {year} {2020})},\ \Eprint
  {http://arxiv.org/abs/2001.08036} {arXiv:2001.08036 [hep-ph]} \BibitemShut
  {NoStop}%
\bibitem [{\citenamefont {Gal}\ \emph {et~al.}(2016)\citenamefont {Gal},
  \citenamefont {Hungerford},\ and\ \citenamefont {Millener}}]{Gal2016}%
  \BibitemOpen
  \bibfield  {author} {\bibinfo {author} {\bibfnamefont {A.}~\bibnamefont
  {Gal}}, \bibinfo {author} {\bibfnamefont {E.~V.}\ \bibnamefont {Hungerford}},
  \ and\ \bibinfo {author} {\bibfnamefont {D.~J.}\ \bibnamefont {Millener}},\
  }\href {\doibase 10.1103/RevModPhys.88.035004} {\bibfield  {journal}
  {\bibinfo  {journal} {Rev. Mod. Phys.}\ }\textbf {\bibinfo {volume} {88}},\
  \bibinfo {pages} {035004} (\bibinfo {year} {2016})},\ \Eprint
  {http://arxiv.org/abs/1605.00557} {arXiv:1605.00557 [nucl-th]} \BibitemShut
  {NoStop}%
\bibitem [{\citenamefont {She}\ \emph {et~al.}(2021)\citenamefont {She},
  \citenamefont {Chen}, \citenamefont {Zhou}, \citenamefont {Zheng},
  \citenamefont {Xie},\ and\ \citenamefont {Xu}}]{She2020}%
  \BibitemOpen
  \bibfield  {author} {\bibinfo {author} {\bibfnamefont {Z.-L.}\ \bibnamefont
  {She}}, \bibinfo {author} {\bibfnamefont {G.}~\bibnamefont {Chen}}, \bibinfo
  {author} {\bibfnamefont {D.-M.}\ \bibnamefont {Zhou}}, \bibinfo {author}
  {\bibfnamefont {L.}~\bibnamefont {Zheng}}, \bibinfo {author} {\bibfnamefont
  {Y.-L.}\ \bibnamefont {Xie}}, \ and\ \bibinfo {author} {\bibfnamefont
  {H.-G.}\ \bibnamefont {Xu}},\ }\href {\doibase 10.1103/PhysRevC.103.014906}
  {\bibfield  {journal} {\bibinfo  {journal} {Phys. Rev. C}\ }\textbf {\bibinfo
  {volume} {103}},\ \bibinfo {pages} {014906} (\bibinfo {year} {2021})},\
  \Eprint {http://arxiv.org/abs/2009.05402} {arXiv:2009.05402 [nucl-th]}
  \BibitemShut {NoStop}%
\bibitem [{\citenamefont {Hempel}\ \emph {et~al.}(2015)\citenamefont {Hempel},
  \citenamefont {Hagel}, \citenamefont {Natowitz}, \citenamefont {R\"opke},\
  and\ \citenamefont {Typel}}]{Hempel2015}%
  \BibitemOpen
  \bibfield  {author} {\bibinfo {author} {\bibfnamefont {M.}~\bibnamefont
  {Hempel}}, \bibinfo {author} {\bibfnamefont {K.}~\bibnamefont {Hagel}},
  \bibinfo {author} {\bibfnamefont {J.}~\bibnamefont {Natowitz}}, \bibinfo
  {author} {\bibfnamefont {G.}~\bibnamefont {R\"opke}}, \ and\ \bibinfo
  {author} {\bibfnamefont {S.}~\bibnamefont {Typel}},\ }\href {\doibase
  10.1103/PhysRevC.91.045805} {\bibfield  {journal} {\bibinfo  {journal} {Phys.
  Rev. C}\ }\textbf {\bibinfo {volume} {91}},\ \bibinfo {pages} {045805}
  (\bibinfo {year} {2015})},\ \Eprint {http://arxiv.org/abs/1503.00518}
  {arXiv:1503.00518 [nucl-th]} \BibitemShut {NoStop}%
\end{thebibliography}

\end{document}